\documentclass{article}
\usepackage{ijcai19}

\usepackage{times}
\usepackage{soul}
\usepackage{url}
\usepackage[hidelinks]{hyperref}
\usepackage[utf8]{inputenc}
\usepackage[small]{caption}
\usepackage{graphicx}
\usepackage{amsmath}
\usepackage{booktabs}
\usepackage[
    backend=biber,
    style=numeric,
    sorting=none
]{biblatex}
\usepackage{makecell}
\usepackage{pifont}
\usepackage{stfloats}
\usepackage{subcaption}
\usepackage{tabularx}

\title{TwoStepDemocracy: Prototyping of self-evolving, democratic, and decentralized systems}

\author{
Stan Verlaan$^1$ and
Johan Pouwelse$^1$\and
\affiliations
$^1$Delft University of Technology, Delft, The Netherlands
}

\begin{document}

\maketitle

\begin{abstract}
Decentralised systems are often built to avoid central control, but their evolution almost always depends on centralised platforms, informal maintainer authority, and a surprising amount of unpaid goodwill. To address this uncomfortable mismatch, we introduce TwoStepDemocracy, a technical proof-of-concept for protocol-native software evolution.

The prototype combines costly cryptographic identities, peer-to-peer dissemination, issue and solution voting, and Bitcoin-based funding campaigns. Users can express demand by proposing and voting on issues; developers can submit concrete solutions; and accepted work can be linked to voluntary, non-custodial funding. The design deliberately separates demand, approval, and payment. This way, money can support a solution, but it never buys more voting power.

The prototype demonstrates that such a coordination layer can be built as a peer-to-peer implementation with local storage, signed governance objects, and Bitcoin integration. We studied performance, scalability, and costs across storage, identity management, and funding. The results show technical feasibility, but not yet social viability. A larger user study is still needed to evaluate whether real communities would, in practice, vote, fund, and coordinate through this mechanism.
\end{abstract}

\section{Introduction}
\label{chapter:introduction}

Software development is entering a dangerous phase of dependency. The tools that once helped communities coordinate code have become the infrastructure through which software itself evolves. GitHub\footnote{\href{https://github.com}{github.com}} and similar platforms no longer merely host repositories. They mediate issue tracking, pull requests, review workflows, continuous integration, security scanning, package distribution, and increasingly AI-assisted development. For many projects, such platforms have become the operating environment of open-source evolution. This is convenient, but it is also fragile. A community that depends on a central platform inherits that platform's outages, pricing decisions, resource limits, and access rules.

\subsection{Centralised forges under AI pressure}
This dependency becomes more problematic in the age of AI-assisted development. GitHub has already described Copilot\footnote{\href{https://github.com/features/copilot}{github.com/features/copilot}} as moving from an in-editor assistant to an agentic platform capable of affecting entire repositories, meaning it can generate commits, open pull requests, run tests, and more at machine speed \cite{GitHub2026CopilotUsageBilling}. The result is not simply faster programming. It creates a resource bottleneck that requires a massive increase in processing power to keep the platform responsive. GitHub itself has acknowledged that recent availability incidents were not acceptable and that the platform is being pushed into a new scaling regime \cite{GitHub2026AvailabilityUpdate}. In the same update, they state that they began planning for a 10$\times$ capacity increase in October 2025, but by February 2026, it became apparent that the future required designing for 30$\times$  today's scale. A few months later, they concluded that the previous pricing model was no longer sustainable. The crisis was so severe that they tightened usage limits and temporarily paused new sign-ups for several premium plans to protect service quality for existing users \cite{GitHub2026CopilotLimits}. These are not signs of a platform that can scale infinitely. They are signs of a scarce computational layer that must ration access. The issue is not that GitHub is uniquely unreliable, but that a centralised development platform forms a shared failure domain. When such a platform degrades, the surrounding software ecosystem degrades with it.

\subsection{The open-source maintenance bottleneck}
This platform pressure falls on an ecosystem already difficult to sustain: the open-source community. Open-source software has become a critical part of modern digital infrastructure, but its long-term development often depends on a small group of maintainers. Empirical studies show that most users remain passive, while the actual development work is performed by a small number of core developers \cite{MockusFieldingHerbsleb2002Oss,AmritHillegersberg2010Coreperiphery}. This concentration of responsibility can be efficient when a project is healthy, but it becomes brittle when the core developer group is unstable. A study of \(1{,}932\) popular GitHub projects found that 16\% had been abandoned. Although 128 of these projects recovered after new core developers took over maintenance, the remaining 187 did not, showing that even popular open-source projects can lose their capacity to evolve \cite{AvelinoConstantinouTulioSerebrenik2019Abandonment}. The benefits of open-source software are widely shared, but the labour of keeping it alive is not.

AI does not automatically solve this imbalance. It may produce more code, but more code is not the same as sustainable evolution. A project still needs to decide which problems matter, which proposed changes to accept, and how to reward useful work. If anything, agentic development makes these questions more urgent. The scarce resource shifts from code generation to coordination.

\subsection{The governance paradox of decentralised systems}
This maintenance bottleneck affects all software projects, but it is even more uncomfortable for decentralised systems. They are designed to avoid central control at runtime, yet their evolution is often coordinated through central platforms. Just as with open source projects, they still require governance mechanisms to determine how protocol updates are proposed, coordinated, and adopted \cite{PeltJansenBaarsOverbeek2020Governance}. Studies of blockchain governance have shown that even systems designed to avoid trusted intermediaries often rely on small groups of core developers to maintain and evolve the protocol \cite{FilippiLoveluck2016Bitcoinpolitics,ArrunadaGaricano2018Blockchain}. The result is a paradox: the protocol may resist central control, while the process that changes the protocol remains centralised or socially opaque.

\subsection{Toward a sovereign software forge}

The previous sections expose three related weaknesses in the current software evolution model of decentralised systems:

\begin{enumerate} 
    \item Centralised forges create shared failure domains.
    \item Open-source maintenance is structurally fragile.
    \item Decentralised protocols evolve through centralised processes.
\end{enumerate}

Together, these weaknesses point to the same missing capability: decentralised software systems need an internal coordination layer. Such a self-supporting development platform should allow users to express demand, developers to respond with implementations, and accepted work to be rewarded within the platform itself. It should not purely mirror GitHub in a peer-to-peer setting. Rather, it should turn software evolution into a protocol-native process without a central coordinator.

The full version of this vision exceeds the scope of this thesis. A complete self-supporting software platform would also need mechanisms for patch distribution, code verification, automatic update delivery, dependency management, and rollback. This work focuses on the coordination layer that must exist before those mechanisms can be meaningful. Specifically, it studies how democratic voting and an incentive market can be combined to support the evolution of a decentralised system.

The contribution of this thesis is a prototype coordination layer for decentralised software evolution. The prototype uses peer-to-peer dissemination for governance objects, cryptographic identities for authenticated participation, and Bitcoin-based funding campaigns for non-custodial developer compensation. The goal is not to remove human judgment from software governance. The goal is to make the basic evolutionary loop explicit: users reveal demand, developers compete to satisfy it, and the community decides which solutions receive legitimacy and payment.

\subsection{Thesis structure}
The rest of this thesis moves from problem diagnosis to construction, and from construction back to reflection. Chapter~\ref{chapter:problem-description} defines the problem and narrows the broader sovereign-forge vision to the part addressed in this work. Chapter~\ref{chapter:background} introduces the required background on Bitcoin transaction primitives, cryptographic identities, and gossip-based dissemination, while Chapter~\ref{chapter:related-work} positions the work against related approaches to decentralised governance, voting, identity, and incentive mechanisms. Chapter~\ref{chapter:design} presents the design of the proposed coordination layer, and Chapter~\ref{chapter:implementation} describes its implementation as a peer-to-peer prototype. Chapter~\ref{chapter:performance} evaluates the prototype's performance and scalability. The final chapters step back from the machinery. Chapter~\ref{chapter:limitations-and-future-work} reflects on limitations and design trade-offs, and outlines future work toward a more complete self-supporting software forge. Finally, Chapter~\ref{chapter:conclusion} concludes the thesis by summarising the main findings.
\section{Problem description}
\label{chapter:problem-description}

The introduction argued that software evolution of decentralised systems is increasingly constrained by centralised forges, fragile maintainer structures, and centralised governance processes. This chapter narrows that broader concern to the specific problem addressed in this thesis. The problem is not simply that decentralised systems need better voting, nor that open-source contributors need better payment. The problem is that there is no integrated coordination mechanism that allows a decentralised system to decide what should change, whether an implementation is accepted, and how useful work should be rewarded.

This thesis does not attempt to build a complete decentralised software forge. In particular, it does not solve automatic deployment, code verification, dependency management, or rollback. These mechanisms are necessary for a complete self-supporting development platform, but they depend on a prior coordination layer. Before a system can safely update itself, it must first decide which problems matter, which solutions are legitimate, and how contributors are incentivised to work on them.

The research problem is therefore the technical design and implementation of this coordination layer. Such a layer must satisfy three core requirements. First, participation must be tied to verifiable identities, so that governance actions cannot be freely duplicated by creating arbitrary accounts. Second, the system must support a workflow from issue discovery to solution acceptance. Third, it must connect accepted work to a funding mechanism without allowing money to directly replace community approval.

This leads to the following research question:

\begin{quote}
    \textit{How can a decentralised protocol support its own evolution through a hybrid mechanism that combines democratic decision-making with public-goods funding, without relying on central maintainers to coordinate issues, approve solutions, and reward contributors?}
\end{quote}

To answer this question, the thesis develops and evaluates a prototype coordination layer. 

\section{Background}
\label{chapter:background}

The proposed system combines peer-to-peer communication, cryptographic authentication, Sybil resistance, and Bitcoin-based payments. This chapter introduces the background concepts needed to understand these mechanisms. Section~\ref{sec:p2p-networks-and-distributed-systems} first introduces peer-to-peer networks and the implications of maintaining only a local view of a distributed system. Section~\ref{sec:gossip-based-dissemination} then explains gossip-based dissemination, which allows protocol objects to spread through the network without a central server. The chapter then turns to identity and participation. Section~\ref{sec:cryptographic-identities-and-signatures} describes how public-key cryptography and digital signatures can be used to authenticate messages and represent pseudonymous identities. Section~\ref{sec:sybil-attacks} explains why such identities are insufficient on their own when influence is assigned per participant. Finally, Section~\ref{sec:bitcoin-transaction-primitives} introduces the Bitcoin transaction primitives used for non-custodial funding, including UTXOs, signature modes, PSBTs, and transaction metadata.

\subsection{Peer-to-peer networks and distributed systems}
\label{sec:p2p-networks-and-distributed-systems}

\begin{figure*}[!b]
    \centering

    \begin{subfigure}[t]{0.32\textwidth}
        \centering
        \begin{minipage}[c][0.28\textheight][c]{\linewidth}
            \centering
            \includegraphics[width=\linewidth, height=0.26\textheight, keepaspectratio]{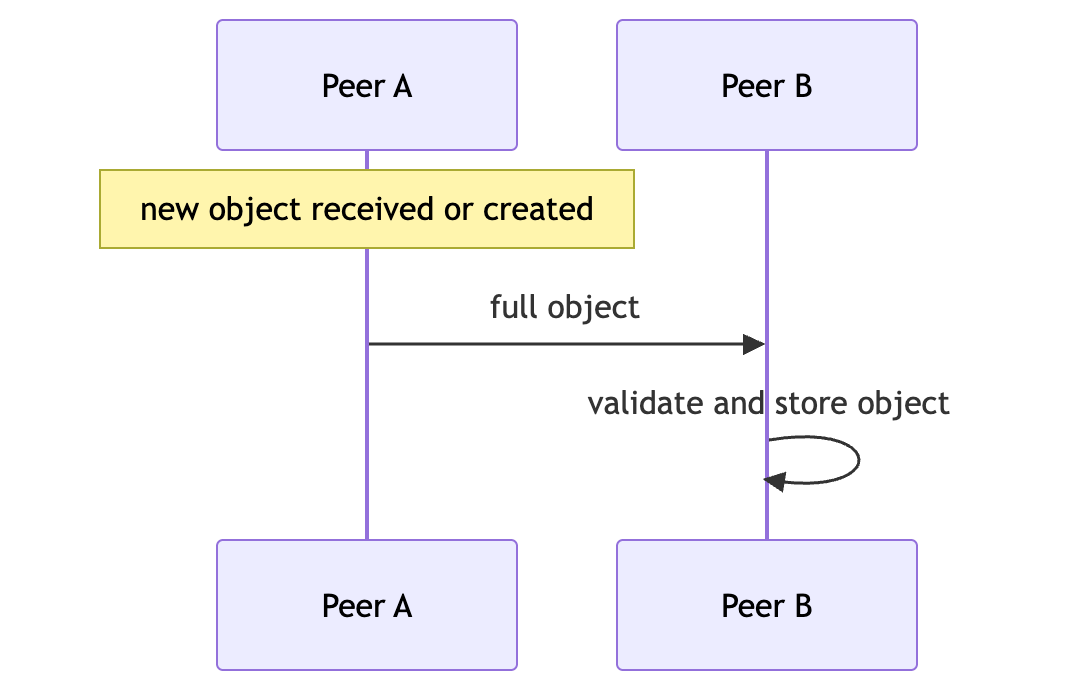}
        \end{minipage}
        \caption{Push}
        \label{fig:gossip-push}
    \end{subfigure}
    \hfill
    \begin{subfigure}[t]{0.32\textwidth}
        \centering
        \begin{minipage}[c][0.28\textheight][c]{\linewidth}
            \centering
            \includegraphics[width=\linewidth, height=0.26\textheight, keepaspectratio]{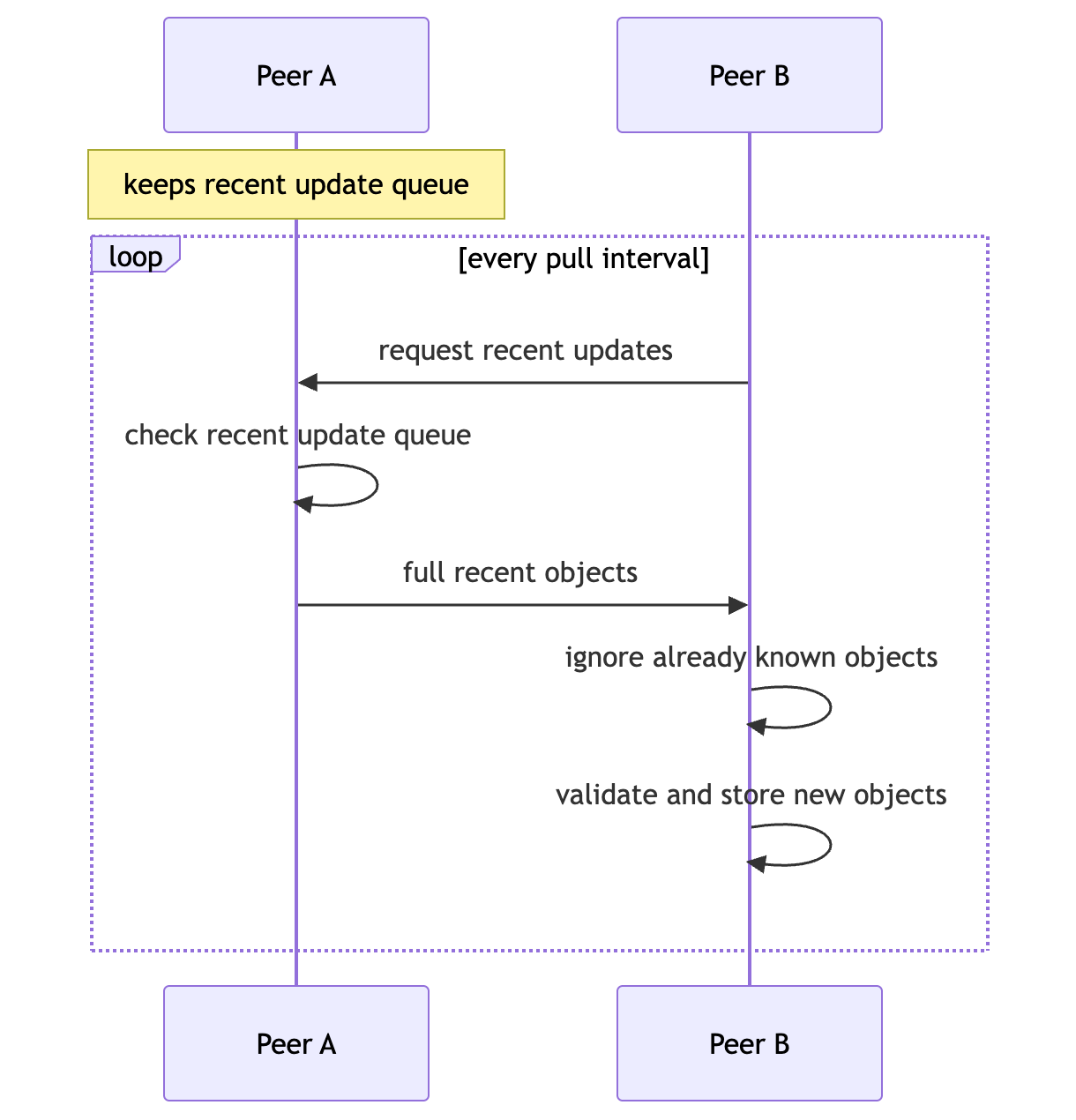}
        \end{minipage}
        \caption{Pull}
        \label{fig:gossip-pull}
    \end{subfigure}
    \hfill
    \begin{subfigure}[t]{0.32\textwidth}
        \centering
        \begin{minipage}[c][0.28\textheight][c]{\linewidth}
            \centering
            \includegraphics[width=\linewidth, height=0.26\textheight, keepaspectratio]{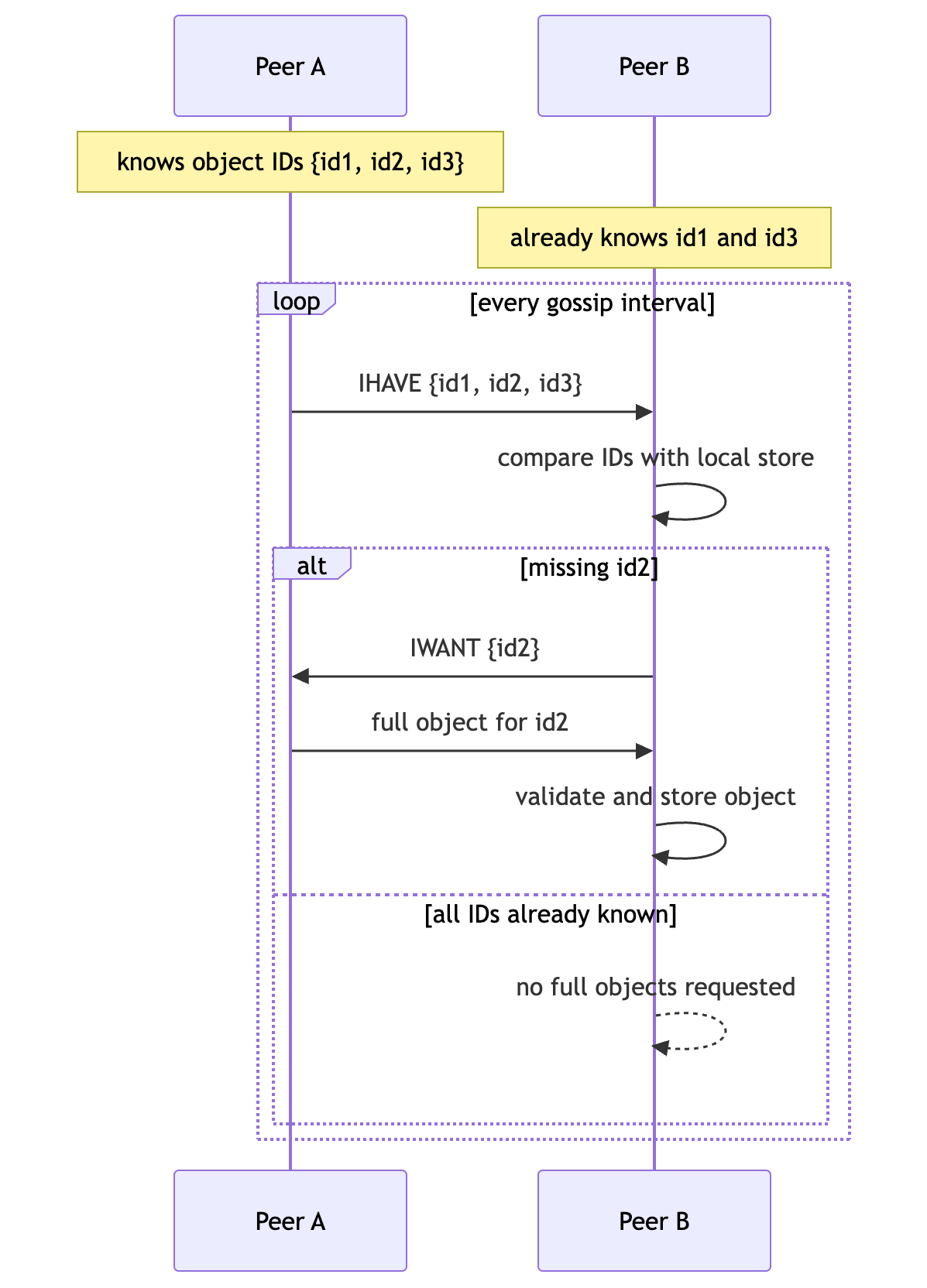}
        \end{minipage}
        \caption{Push-pull}
        \label{fig:gossip-push-pull}
    \end{subfigure}

    \caption{Comparison of push, pull, and push-pull dissemination. In push dissemination, full objects are sent immediately upon creation or receipt. In pull dissemination, peers periodically ask whether new objects are available. In push-pull dissemination, peers first advertise compact object identifiers, after which receivers request only missing objects.}
    \label{fig:gossip-dissemination-comparison}
\end{figure*}

A distributed system consists of multiple independent nodes that communicate over a network to provide shared functionality. In a centralised architecture, clients depend on a server that stores or coordinates the relevant state. In a peer-to-peer (P2P) architecture, participants communicate directly with each other and can both receive and forward information. This reduces dependence on a single operator, but it also means that no peer necessarily observes the complete system state \cite{Schollmeier2001PeerToPeer,RFC5694PeerToPeer}.

This has important consequences for protocol design. Messages may be delayed, lost, duplicated, or received in different orders by different peers. As a result, each peer maintains only a local view of the system. The absence of an object in one peer's store does not prove that the object does not exist, as it may simply not have been received yet.

\subsection{Gossip-based dissemination}
\label{sec:gossip-based-dissemination}

Gossip-based dissemination is a family of decentralised communication protocols in which information spreads through repeated message forwarding between peers. Instead of relying on a central server or requiring one node to broadcast directly to every other node, each peer forwards information to a subset of its neighbours. Over time, this causes information to propagate through the network in an epidemic manner \cite{Demers1987Epidemic}. Gossip protocols are therefore commonly used in distributed systems where robustness, decentralisation, and tolerance to churn are more important than immediate global consistency.

A basic distinction can be made between push, pull, and push-pull dissemination, as illustrated in Figure~\ref{fig:gossip-dissemination-comparison}. In a push protocol, informed peers actively send information to other peers, as shown in Figure~\ref{fig:gossip-push}. In a pull protocol, uninformed peers request information from peers that may already have it, as shown in Figure~\ref{fig:gossip-pull}. Push-pull protocols combine both mechanisms: peers announce what information they have, while receivers request the information they are missing. This can reduce redundant communication compared to naive flooding, where complete messages are repeatedly sent even to peers that already received them \cite{KarpSchindelhauerShenkerVocking2000RumorSpreading}.

Modern P2P push-pull protocols often apply this idea by separating metadata dissemination from full message transfer. For example, libp2p's GossipSub uses control messages such as \texttt{IHAVE} and \texttt{IWANT}. An \texttt{IHAVE} message advertises message identifiers known by a peer, while an \texttt{IWANT} message requests the full contents of unknown messages \cite{VyzovitisNaporaMcCormickDiasPsaras2020GossipSub}. This allows peers to first compare compact identifiers and only transfer complete payloads when needed. The push-pull exchange in Figure~\ref{fig:gossip-push-pull} shows this pattern at a high level. Later versions of GossipSub also include measures to limit spam and resource exhaustion, such as capping \texttt{IHAVE} advertisements and limiting repeated responses to \texttt{IWANT} requests \cite{Libp2pGossipsubV11}.

Because gossip protocols disseminate information asynchronously, different peers may temporarily observe different local views of the network. Gossip-based systems, therefore, usually provide eventual dissemination rather than immediate global consistency. This trade-off is acceptable in many P2P systems because it avoids central coordination while still allowing independently verifiable messages to spread through the network.

\subsection{Cryptographic identities and signatures}
\label{sec:cryptographic-identities-and-signatures}

Public-key cryptography enables an entity to generate a key pair comprising a private key and a corresponding public key. The private key is kept secret, while the public key can be distributed openly. In decentralised systems, a public key can act as a pseudonymous cryptographic identity: it does not reveal the real-world identity of the holder, but it allows actions to be linked to the same key pair.

Digital signature schemes allow messages to be authenticated using such key pairs. A signature is created with the private key and verified with the corresponding public key. A valid signature provides authenticity, because it shows that the message was authorised by the holder of the private key, and integrity, because modifying the signed message invalidates the signature \cite{NIST2023DSS}.

Signatures should also be bound to a specific context. Protocols often include fields such as a protocol label, version number, network identifier, or action type in the signed message. This technique, known as domain separation, reduces the risk that a signature created for one purpose is accepted in another context \cite{RFC9380HashToCurve}.

Authentication protocols must also ensure freshness. A valid signature may otherwise be copied and reused in a replay attack. Challenge-response protocols address this by requiring the signer to sign a fresh challenge, often containing a nonce. Since the nonce is intended to be used only once, the verifier can distinguish a new authentication attempt from a reused old message \cite{NIST2025DigitalIdentity}.

Cryptographic identities are useful because they allow peers to verify messages without relying on a central account provider. However, they do not provide uniqueness on their own: a single actor can generate many key pairs at negligible cost. Systems that assign influence, trust, or voting power per identity, therefore, need an additional mechanism to prevent skewed influence from one person with many identities.

\subsection{Sybil attacks}
\label{sec:sybil-attacks}

This identity-multiplication problem is known as a Sybil attack. The term was introduced for attacks in which one entity presents itself as many independent participants in a distributed system \cite{Douceur2002Sybil}. The attack is especially relevant in open systems, where participants can join freely and where protocol decisions depend on the assumption that different identities represent sufficiently independent actors.

Sybil resistance refers to mechanisms that make such identity multiplication harder or more expensive. Different systems approach this through central identity providers, social trust relationships, proof-of-work, proof-of-stake, or other resource-based costs. These mechanisms do not all provide the same guarantees, but they address the same underlying problem: participation in an open system should not be cheaply duplicated.

\subsection{Bitcoin transaction primitives}
\label{sec:bitcoin-transaction-primitives}

\begin{figure}[htbp]
    \centering
    \includegraphics[width=\linewidth]{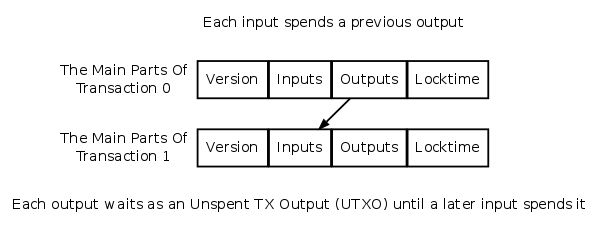}
    \caption{Bitcoin transactions consume outputs from earlier transactions and create new outputs that can later be spent. Reproduced from the \href{https://developer.bitcoin.org/devguide/transactions.html}{Bitcoin Developer Documentation}, licensed under the \href{https://opensource.org/licenses/MIT}{MIT License}.}
    \label{fig:bitcoin-inputs-outputs}
\end{figure}

Bitcoin uses an unspent transaction output (UTXO) model. In this model, value is not represented as an account balance that is updated over time. Instead, Bitcoin transactions consume outputs from earlier transactions and create new outputs that can be spent later, as shown in Figure~\ref{fig:bitcoin-inputs-outputs}. Each spendable output is identified by the transaction that created it and by its position within that transaction, commonly written as \texttt{txid:vout}. A wallet balance is therefore the sum of all UTXOs controlled by that wallet. Spending Bitcoin means constructing a transaction that consumes one or more UTXOs as inputs and assigns their value to one or more new outputs \cite{BitcoinDeveloper2025Transactions}.

This model is important because transaction inputs can be treated as independent pieces. A user can choose a specific UTXO and authorise it to be spent under particular transaction conditions. The authorisation is expressed through a digital signature. In the default case, a Bitcoin signature commits to the transaction in such a way that changing the transaction invalidates the signature. However, Bitcoin supports different signature hash modes, which determine which parts of the transaction are covered by the signature.

\subsubsection{Signature modes}
One relevant signature mode is \texttt{SIGHASH\_ALL}. This mode commits the signer to all transaction outputs. If any output address or amount is changed, the signature becomes invalid. This protects the signer from having their input redirected to a different recipient. Another relevant flag is \texttt{SIGHASH\_ANYONECANPAY}. When this flag is used, the signer only commits to their own input, while allowing other inputs to be added later. When combined as \texttt{SIGHASH\_ALL | SIGHASH\_ANYONECANPAY}, the signer authorises their own input for a transaction with fixed outputs, while leaving room for other participants to contribute additional inputs.

\subsubsection{Partially signed Bitcoin transactions}
Bitcoin transactions can also be prepared and signed in multiple steps. Partially Signed Bitcoin Transactions (PSBTs) provide a standard format for this workflow. A PSBT contains an unsigned or partially signed transaction together with the metadata needed by wallets and signing devices. This separates transaction construction from transaction signing. One party can draft the transaction template, while the other signs it with their own wallet. This is useful in systems where the application should not hold the user's private keys and where signing may happen in external wallet software \cite{BIP2017PSBT}.

\subsubsection{Metadata in output}
A transaction may also contain an \texttt{OP\_RETURN} output. Such an output is provably unspendable and can carry a small amount of arbitrary data. Because the output cannot be spent, it cannot be used to transfer value. Instead, it is often used to commit metadata to the blockchain. A common pattern is to store only a hash of application-level data, rather than the data itself. This allows external systems to later verify that a transaction corresponds to a specific off-chain object or agreement, while keeping the on-chain footprint small.

These primitives are useful for protocols that need to coordinate Bitcoin payments without giving custody of funds to an intermediary. UTXOs identify the exact coins that may be spent, PSBTs allow transaction signing to happen outside the application, signature hash modes control which parts of a transaction remain fixed, and \texttt{OP\_RETURN} commitments allow transactions to be linked to off-chain protocol objects. Together, these mechanisms enable the design of payment flows in which users retain control over their funds until a valid transaction is eventually broadcast.

\section{Related work}
\label{chapter:related-work}

\newcommand{\cmark}{\ding{51}}
\newcommand{\xmark}{\ding{55}}
\newcommand{\pmark}{$\sim$}

\begin{table*}[!b]
    \centering
    \caption{Comparison of governance approaches against the coordination requirements addressed in this thesis.}
    \label{tab:governance-approach-comparison}
    \scriptsize
    \renewcommand{\arraystretch}{1.2}
    \begin{tabularx}{\textwidth}{
        >{\raggedright\arraybackslash}X
        c
        c
        c
        c
    }
        \toprule
        \textbf{Approach} &
        \makecell{\textbf{Issue prioritisation}} &
        \makecell{\textbf{Solution acceptance}} &
        \makecell{\textbf{Contributor compensation}} &
        \makecell{\textbf{Voting separated}\\\textbf{from funding}} \\
        \midrule
        Off-chain rough consensus &
        \pmark &
        \pmark &
        \xmark &
        \cmark \\
        On-chain token-weighted governance &
        \pmark &
        \cmark &
        \xmark &
        \xmark \\
        Delegated or liquid democracy &
        \xmark &
        \pmark &
        \xmark &
        \cmark \\
        TwoStepDemocracy &
        \cmark &
        \cmark &
        \cmark &
        \cmark \\
        \bottomrule
    \end{tabularx}

    \vspace{0.5em}
    \footnotesize{\cmark = supported; \pmark = partially supported or handled informally; \xmark = not directly supported.}
\end{table*}

This chapter positions the proposed system in relation to existing work on decentralised governance and the mechanisms needed to support it. The first section discusses approaches that address the overarching problem of protocol evolution. These systems show how decentralised communities can coordinate change, but also expose recurring tensions between participation, authority, and economic power. The remaining sections focus on the individual mechanisms required by such a governance system. Section~\ref{sec:identity-in-decentralised-systems} examines identity and Sybil resistance, since governance decisions only become meaningful when influence cannot be gained cheaply. Section~\ref{sec:public-goods-funding-bounties-and-crowdfunding} discusses different ways to connect useful work to compensation. Together, these areas form the design space for the coordination layer proposed in this thesis.

\subsection{Decentralised governance and protocol evolution}
\label{sec:decentralised-governance-and-protocol-evolution}

Several approaches have been proposed or adopted to govern the evolution of decentralised systems. A first approach is off-chain governance through proposals, discussion, and rough consensus. Bitcoin Improvement Proposals (BIPs), for example, provide a structured process for proposing changes to Bitcoin, while acceptance still depends on discussion, implementation, review, and eventual adoption by the ecosystem \cite{BIP3}. This model avoids placing all governance logic within the protocol, giving developers and users room to evaluate changes socially. Its limitation is that the decision process remains partly informal since discussion, review, and adoption are coordinated outside the protocol, and the boundary between technical judgment and social authority is not always explicit.

A second approach is on-chain governance, where protocol evolution is made explicit inside the system itself. Tezos is a prominent example of a blockchain with an on-chain self-amendment mechanism. Protocol changes can be proposed, voted on, tested, and activated through a built-in governance process, reducing the need for disruptive hard forks \cite{TezosGovernance}. This makes protocol evolution more explicit than purely off-chain governance. Similar ideas also appear in decentralised autonomous organisations (DAOs), where members can vote on proposals if they own governance tokens. These systems provide a clear and deterministic mechanism. However, they usually tie voting power to stake or token ownership. As a result, power remains tied to economic ownership of the underlying system. Empirical studies of DAO governance prove that voting power is often unevenly distributed, leading to substantial influence by a small number of actors over proposal outcomes \cite{FritschMullerWattenhofer2022DAOControl}. On-chain governance, therefore, makes decision-making more explicit, but it does not necessarily provide democratic equality among participants.

Delegated or liquid democracy has been proposed as a way to address low participation and expertise problems in decentralised governance. In a liquid democracy, participants may either vote directly or delegate their voting power to another participant. Delegation can improve participation when users are inactive or lack expertise, and it allows influence to flow toward more trusted or knowledgeable participants. However, liquid democracy introduces risks, such as delegation cycles and the concentration of delegated power \cite{ChristoffGrossi2017LiquidDemocracy}. Studies of delegated voting in blockchain systems show that it can increase participation, but it may also lead to power concentration through delegation networks \cite{LiXuDuan2023LiquidDPoS}.

Table~\ref{tab:governance-approach-comparison} summarises how these approaches relate to the coordination requirements addressed in this thesis. The design of TwoStepDemocracy separates issue prioritisation, solution acceptance, and developer compensation. Registered participants indicate which issues matter, vote separately on concrete solutions, and voluntarily fund accepted work. The novelty of TwoStepDemocracy is its explicit coordination layer, which also integrates a funding mechanism.

\subsection{Identity in decentralised systems}
\label{sec:identity-in-decentralised-systems}

The previous section showed that governance mechanisms require participants to perform actions that can later be attributed, verified, and counted. This makes identity a central design problem. If governance power is assigned per participant, the system needs a way to prevent a single actor from cheaply appearing as multiple participants. Otherwise, elections can be distorted by duplicated identities.

Sybil resistance is therefore an important consideration in the design. Cryptocurrencies often do not prevent users from creating multiple accounts directly. Instead, they limit the influence of identities through economic mechanisms. Bitcoin, for example, uses Proof of Work, where influence in the consensus process depends on computational effort rather than the number of accounts \cite{Nakamoto2008Bitcoin}. This provides a probabilistic security guarantee as long as the majority of computational power follows the protocol. Ethereum addresses the same problem differently through Proof of Stake, where influence depends on staked capital, and malicious behaviour can be financially penalised \cite{Ethereum2026Pos}.

This approach works well in systems where influence can be tied to scarce resources, such as computation or capital. Voting systems, however, do not always have such a resource and usually require each participant to be represented by exactly one account and therefore to cast only one vote per election. In close elections, a single participant casting multiple votes may already be sufficient to change the result.

Several systems have therefore proposed or implemented mechanisms to strengthen identity guarantees. One example is SmartPhoneDemocracy, a proposal for a decentralised voting system that uses the European Commission's EUDI wallet for identity management \cite{JozwikPouwelse2025SmartphoneDemocracy}. This approach has clear advantages, as it can provide strong guarantees that each participant corresponds to a real person. However, it also introduces limitations. The EUDI wallet is still under development and not yet fully deployed; it is only applicable to EU citizens, and identities are issued by recognised authorities \cite{Europeancommission2026Eudi}. As a result, the system relies on a centralised identity infrastructure and does not provide the same level of pseudonymity as systems that allow users to create their own cryptographic identities.

Gitcoin is another project that has used voting rounds as part of its governance and funding mechanisms. To improve Sybil resistance, Gitcoin Passport (now Human Passport) links accounts to external verifiable credentials, such as web2 activity, web3 activity, biometrics, or web-of-trust signals \cite{Humanpassport2026Docs}. This approach raises the cost of creating many convincing identities, but it also depends on existing external platforms and the trust assumptions associated with them.

Another approach is the use of Proof of Personhood mechanisms, although these have not yet been adopted on a large scale. Such systems grant accounts to users who can provide evidence that they correspond to a unique physical person. In some cases, this requires being introduced by already verified members, while other systems require users to attend an in-person verification event \cite{BorgeKokorisJovanovicGasserGaillyFord2017Pop,SiddarthIvlievSiriBerman2020Watchmen}. These approaches can provide stronger identity guarantees, but they also make onboarding more difficult. As a result, growth is often slow, since verifying new users requires time, coordination, and active participation from the existing network.

The identity mechanism used in this thesis takes a different position in this design space. It uses a registration payment. An account becomes valid only after an application-level public key is linked to a Bitcoin payment. This does not prove that each account corresponds to a unique human participant. However, it does make identity creation economically costly, which raises the cost of large-scale Sybil attacks while preserving pseudonymity and avoiding the onboarding burden of stronger identity systems.

\subsection{Public goods funding, crowdfunding, and bounties}
\label{sec:public-goods-funding-bounties-and-crowdfunding}

The motivation for TwoStepDemocracy is that decentralised systems require continuous development. Bugs need to be fixed, features need to be implemented, and protocols need to adapt over time. If such work relies only on voluntary contributions, important improvements may remain unimplemented, especially when they are useful to the community but costly for individual developers to perform. This creates a need for mechanisms that connect development work to compensation.

One approach is philanthropic or community-based matching. Quadratic funding is a prominent mechanism in this category. In quadratic funding, users make individual contributions to projects, and an additional pool of shared funds is distributed to them based on the breadth of support they receive. Projects supported by many distinct contributors receive proportionally more matching funds than projects supported by only a few large contributors \cite{ButerinHitzigWeyl2019LiberalRadicalism}. This mechanism is used in ecosystems such as Gitcoin Grants to fund public goods based on broad community support rather than solely on the total amount donated \cite{Gitcoin2026QuadraticFunding}. Quadratic funding is useful when a community has a shared budget to allocate across multiple projects.

A second approach is crowdfunding through threshold mechanisms. In an assurance contract, contributors pledge funds toward a public good, but the project is funded only if the total contributions reach a predefined threshold. If the threshold is not reached, contributors are refunded \cite{BagnoliLipman1989ProvisionPublicGoods}. Blockchain systems can implement this logic through smart contracts, where the threshold condition and fund release are enforced by the contract rather than by a central crowdfunding platform. LikeStarter, for example, proposes an Ethereum-based decentralised autonomous organisation for crowdfunding without a central authority \cite{ZichichiContuFerrettiDAngelo2019LikeStarter}. Such systems reduce dependence on an intermediary and make the campaign state transparent.

A third approach is the use of bounties. In a bounty system, users attach a monetary reward to an issue, and a developer receives it upon completing the task. Empirical work on one such system, BountySource, shows that bounties for GitHub projects can help attract developer attention to issues, especially when proposed early and with a value high enough for the project context \cite{Zhou2021BountiesGitHub}. At the same time, bounties do not eliminate the need to judge whether the work actually solves the issue. In traditional open-source projects, this judgment is usually made by maintainers, who decide whether to accept a pull request.

The funding campaigns in TwoStepDemocracy combine elements of these approaches. A campaign has a funding target, but it is linked to a submitted solution rather than to a project or open issue. Funding and acceptance are also separated \cite{HalesRahmanZhangMeulpolderPouwelse2009BitCrunch}. Users may pledge financial support, but the solution is accepted only if it passes a separate solution vote. This keeps willingness to pay distinct from community approval while still providing developers with concrete economic compensation.

\section{Design}
\label{chapter:design}

\begin{table*}[!b]
	\centering
	\caption{Overview of the main protocol objects.}
	\label{tab:protocol-objects}
	\begin{tabular}{p{0.24\linewidth} | p{0.68\linewidth}}
		\textbf{Object} & \textbf{Description} \\
		\hline
		\texttt{Registration} & Links an application-level public key to a Bitcoin registration transaction. \\
		\hline
		\texttt{Issue} & Describes a bug, feature request, or improvement. It contains the problem description and metadata that users and developers need to understand the request. \\
		\hline
		\texttt{IssueVote} & Records support for an issue by a registered participant. Each participant can cast at most one valid vote per issue. \\
		\hline
		\texttt{Solution} & Describes a concrete implementation submitted for an issue. It is linked to the issue it addresses. \\
		\hline
		\texttt{SolutionVote} & Records approval for a specific submitted solution. Each participant can cast at most one valid vote per solution. \\
		\hline
		\texttt{FundingCampaign} & Defines the payment terms for a solution. It links to a solution and fixes the asking price, payout address, and deadline. \\
		\hline
		\texttt{FundingPledge} & Represents a contributor's signed commitment to fund a campaign with a specific Bitcoin UTXO. \\
		\hline
		\texttt{FundingTransaction} & Confirmed Bitcoin transaction that combines valid pledges and pays the developer. It completes the funding campaign. \\
	\end{tabular}
\end{table*}

From the previous chapters, we can conclude that decentralised systems need a way to evolve without depending entirely on a fixed group of core maintainers. Users must be able to express which problems they consider important, while developers need clear signals about which work is worth doing, and they should be compensated for it. The design presented in this section introduces a protocol-native coordination mechanism that does exactly those things.

At a high level, the protocol follows the lifecycle of an issue. A registered user creates an issue; users vote on it to express demand; after which, a developer may submit a concrete solution. This submission opens two parallel processes: a solution vote and a funding campaign. A solution is successful only when it receives sufficient user approval, and the developer's asking price is matched.

The central design principle is simplicity. Many governance mechanisms become difficult to adopt because they try to solve too many problems at once. They mix voting with capital, delegate legitimacy to complex identity systems, or introduce decision rules that users may struggle to understand and trust. This design instead chooses a narrower path. Identity is made costly through a registration payment, rather than through a full proof-of-personhood system. Governance is split into issue prioritisation and solution acceptance, rather than being hidden inside an informal maintainer's judgment. Compensation is handled through funding campaigns that are tied to concrete submitted work, but kept separate from the vote that decides whether the work is accepted. The novelty is therefore not an isolated mechanism, but the way these simple mechanisms are composed into a protocol workflow in which problems, solutions, approval, and payment are all explicit.

The rest of the chapter dives deeper into the design. Section~\ref{sec:protocol-objects} first introduces the protocol objects that form the system's basic data model. The following section describes the governance process built from these objects. Section~\ref{sec:identity} then presents the identity mechanism. Section~\ref{sec:message-dissemination} explains how valid objects are exchanged between peers using gossip-based dissemination. The funding campaign section then describes how submitted solutions are linked to non-custodial Bitcoin pledges and final settlement transactions. Finally, the chapter discusses the main limitations of the design.

\subsection{Protocol objects}
\label{sec:protocol-objects}

TwoStepDemocracy is built around a small set of immutable objects that are created, signed, verified, stored, and disseminated by peers. Each object represents one action or state transition in the governance process. Peers derive their local protocol state from the set of valid objects they know, rather than from a central database. Table~\ref{tab:protocol-objects} gives an overview of the main protocol objects and their role in the system.

These objects are append-only: once an object has been created and signed, it is not modified. Peers derive local state by combining valid objects rather than by mutating a central database. For example, vote counts and funding totals are computed from known votes, pledges, campaigns, and confirmed Bitcoin transactions. This fits the gossip-based dissemination model described later in section \ref{sec:message-dissemination}.

\begin{figure*}[t]
    \centering
    \includegraphics[width=\textwidth]{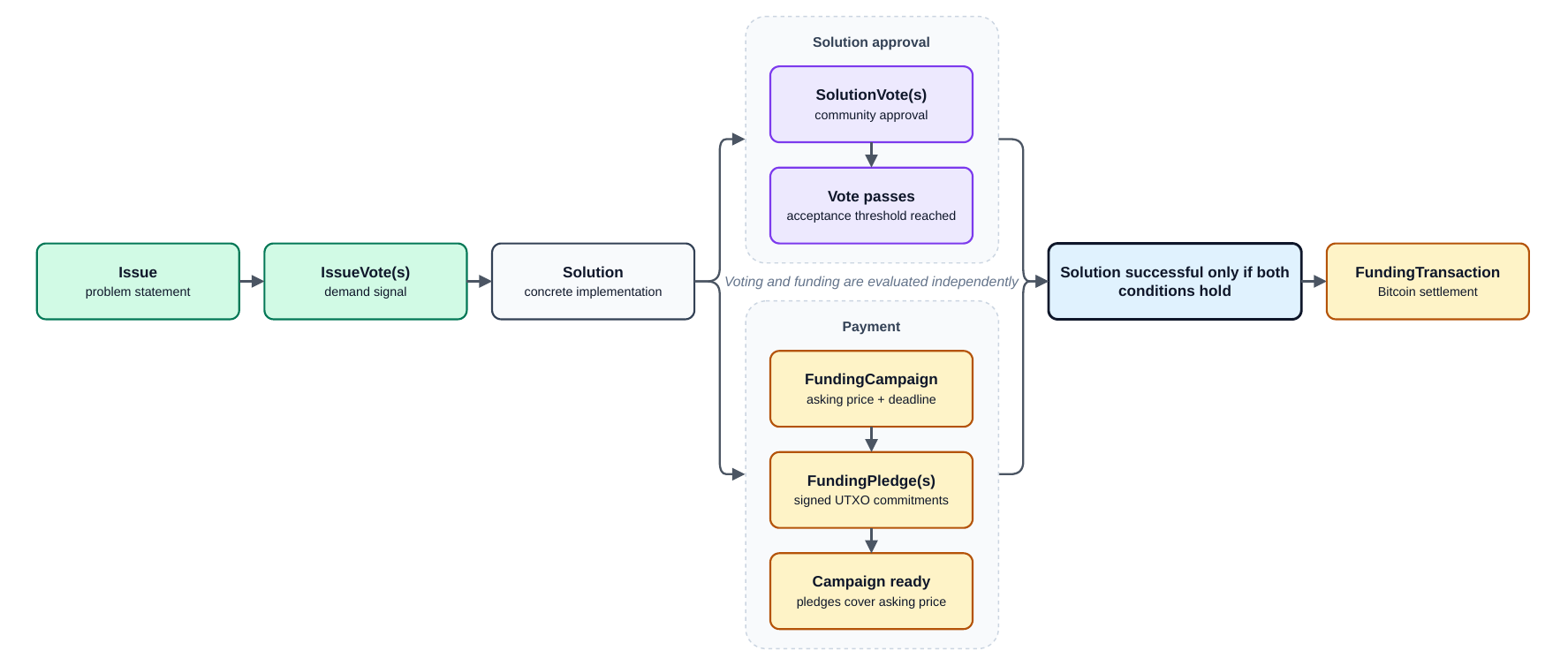}
    \caption{Overview of the TwoStepDemocracy governance process. Users first express demand through issue votes. A submitted solution then opens two independent conditions: community approval through solution votes and economic support through a funding campaign. A solution is successful only when both conditions are satisfied.}
    \label{fig:governance process}
\end{figure*}

\subsection{Governance process}
\label{sec:governance-process}

The governance process is based on a separation between demand, approval, and payment. These concerns are related, but they answer different questions. Issue voting measures whether users consider a problem important. Solution voting determines whether users accept a specific implementation of that problem. Funding determines whether the community is also willing to compensate the developer.

The first voting phase is only used for preference aggregation. A vote on an issue expresses support for the problem statement, not for any implementation yet. This is a deliberate middle ground. If issue prioritisation is kept entirely outside the protocol, developers receive only informal and fragmented signals about what the community wants. If it is made too binding, however, the system risks locking funds for a problem before any concrete solution exists. The issue vote is therefore treated as a public demand signal for developers. It helps developers decide which problems are worth working on, but it does not create an obligation to address the issue or reserve funds.

The second voting phase is about a submitted solution rather than the original issue. At this point, users no longer vote on whether the problem is relevant, but on whether this particular implementation should become part of the system. Each registered user can vote for a solution at most once. A solution passes the voting condition only if it reaches the required acceptance threshold. This makes the solution vote a separate confirmation step: even if an issue was popular, a proposed implementation must still be accepted on its own merits.

Funding is separated from the solution vote. When submitting a solution, the developer specifies an asking price and a funding campaign is opened for that solution. Users may contribute to this campaign voluntarily, but contributions do not increase voting power. Likewise, voting in favour of a solution does not require a financial contribution. This prevents the acceptance rule from becoming a direct pay-to-vote mechanism, while still allowing users to financially support implementations they want to see completed.

A solution is successful only when both conditions are satisfied: the solution vote must pass, and the linked funding campaign must be funded. The protocol, therefore, requires both social approval and economic support before a submitted solution can be considered complete. A full view of the system is shown in Figure~\ref{fig:governance process}.

\subsection{Identity}
\label{sec:identity}

The governance mechanism proposed in the previous section requires participants to vote, submit issues, propose solutions, and interact with other peers. For these actions to be meaningful, the system needs some notion of participant identity. However, this does not mean that the system needs to establish strong legal identity or full proof of personhood. The elections considered here are not state elections or safety-critical decisions over scarce public resources. They are coordination mechanisms for prioritising issues and accepting software solutions. The identity mechanism should therefore be strong enough to discourage cheap duplication, but not so heavy that participation depends on legal identity providers, in-person ceremonies, or complex external credential systems.

This design uses economically costly cryptographic identities. A participant is represented by an application-level public key, linked to a Bitcoin transaction that serves as proof of participation in the system. This is weaker than proof of personhood since it does not prove that each account corresponds to a unique human being. However, unlike a free key pair, it makes each account cost money and expire after a defined period. This raises the cost of Sybil attacks while preserving pseudonymity and keeping onboarding simple enough for a decentralised setting.

The choice is also motivated by the absence of a natural scarce resource in the governance process. In Bitcoin's consensus mechanism, influence is tied to proof-of-work; in proof-of-stake systems, it is tied to locked capital. In this system, issue and solution votes should not be weighted by money, because funding and approval are intentionally separated. A registration payment is therefore used only as a barrier to identity multiplication, not as a source of voting power.

The registration payment is made to an address that we call the treasury address. The use of a treasury address stems from a broader system vision outside the scope of this thesis. The identity mechanism was originally designed for a larger system, Superorganism, in which TwoStepDemocracy would be combined with MyCelium, a self-replicating seedbox infrastructure \cite{DogariuPouwelse2026SelfReplicatingSeedboxes}. In that setting, registration payments could fund a treasury controlled by the seedboxes and later be used to acquire or maintain additional infrastructure. This thesis does not implement any of that. Within TwoStepDemocracy, the payment is used only as evidence of costly registration; the collected funds are not spent by the protocol.

The main design problem becomes a linkage problem. A valid participant should be able to prove that a Bitcoin transaction exists, that the transaction was intended as a registration payment for this system, and that the participant controls the private key corresponding to the registered public key. Later, when the participant votes or communicates with peers, these actions must be linked back to the same registered identity without allowing replay or credential hijacking. The identity design therefore links four elements: the Bitcoin payment, the application public key, the authenticated peer session, and the governance actions performed by that participant.

\subsubsection{Registration}

The first thing a participant must do is create an account. During registration, a user creates an application-level key pair. The public key becomes the user's identity within the system. The user then makes a payment to the community's treasury address. The resulting transaction ID is used as the participant's proof of payment.

A transaction ID alone is not sufficient, because any user could point to a payment made by someone else. For this reason, the payment must be bound to the user's public key. We do this by including a commitment to the public key in the transaction metadata. In the current design, this commitment is computed as:
\[
    H(\texttt{protocol version} \mid \texttt{network} \mid \texttt{public key})
\] \\

where $H$ is a cryptographic hash function, \texttt{network} distinguishes between environments such as regtest, testnet, and mainnet, and \texttt{public key} is the application-level public key. Including the protocol version and network prevents the same commitment from being reused unintentionally across incompatible versions or networks.

To verify a registration, the system checks that the transaction exists, that it pays to the correct treasury address, that it satisfies the required payment policy, and that the metadata contains the expected commitment. If these checks succeed, the transaction and public key together define a registered account. This does not prove that the user corresponds to a unique human being, but it does make each additional identity economically costly.

\subsubsection{Authentication}

Registration proves that a public key was linked to a payment, but it does not by itself prove that the current user controls the corresponding private key. Therefore, authentication requires a second step. During login or peer interaction, the user must sign a fresh commitment using the private key associated with the registered public key.

The commitment includes contextual information such as the protocol name, network, action, account ID, nonce, and issue time. A human-readable version of the signed message is shown below:

\begin{verbatim}
protocol    = superorganism-login-v1
network     = {regtest|testnet|mainnet}
action      = login
public_key  = {public key}
nonce       = {nonce}
issued_at   = {issued at}
\end{verbatim}

The nonce ensures freshness, while the issue time limits the period in which the signature can be reused. This prevents an attacker from authenticating with only copied public account data, such as a transaction ID and public key. A valid authentication attempt must therefore show both that the registration payment exists and that the user controls the private key associated with it.

\subsubsection{Communication}

Because the system is peer-to-peer, authentication should not only happen during local login. Peers should also verify that the nodes they communicate with are registered participants. Otherwise, unauthorised nodes could consume bandwidth, relay invalid data, or attempt to influence the governance process without contributing to the system.

\begin{figure}[htbp]
    \centering
    \includegraphics[width=\linewidth]{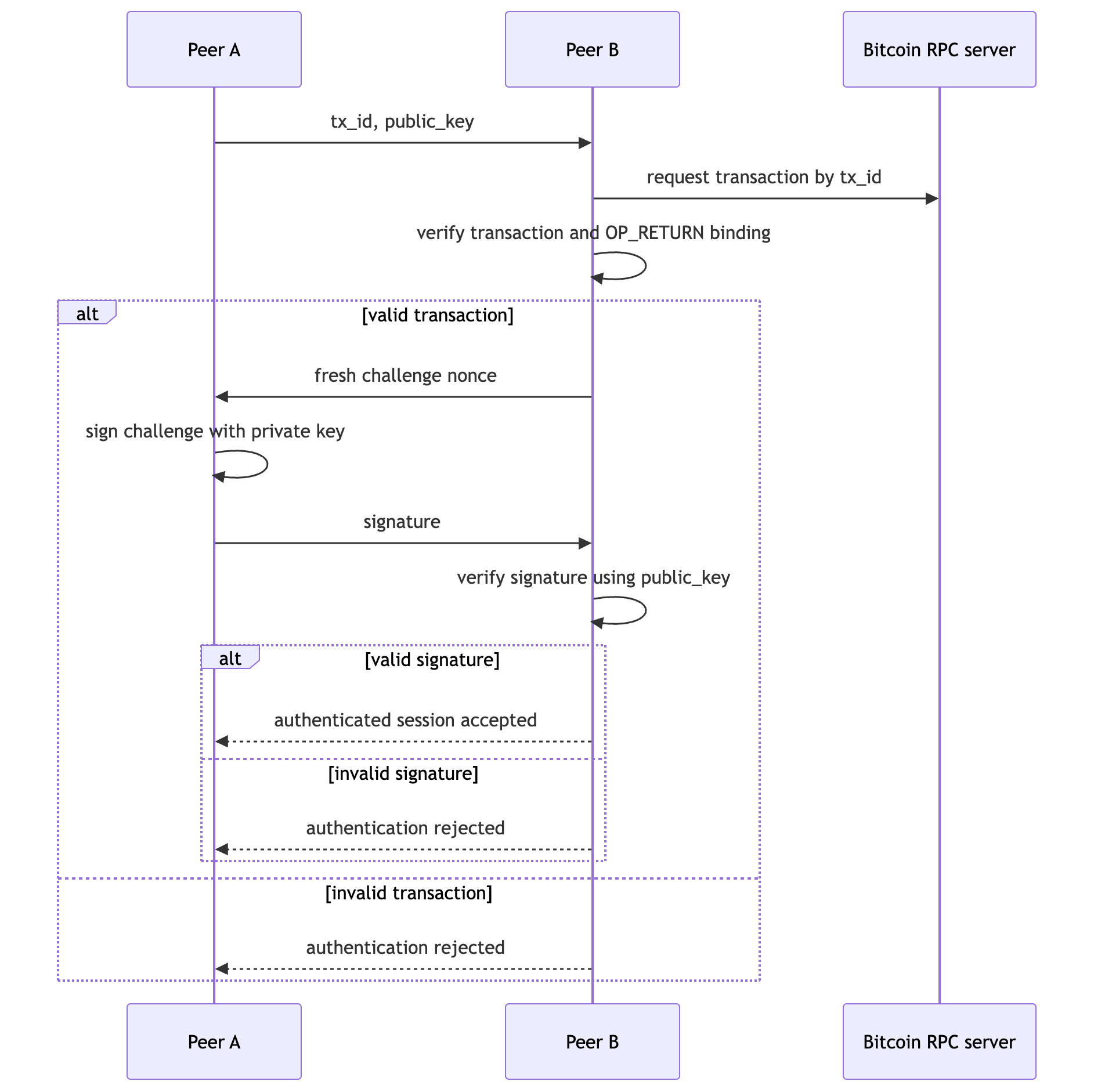}
    \caption{Peer authentication flow linking a Bitcoin registration transaction to an application-level public key and an authenticated peer session.}
    \label{fig:peer-authentication-flow}
\end{figure}

The communication flow follows a challenge-response pattern. When two peers connect, the initiating peer first sends its transaction ID and public key. The receiving peer verifies that the transaction is a valid registration payment and that it is bound to the provided public key. The receiving peer then sends a fresh challenge. The initiating peer signs this challenge using its private key and returns the signature. If the signature is valid, the receiving peer temporarily accepts the remote peer as the owner of that registered identity. The complete flow can be seen in Figure~\ref{fig:peer-authentication-flow}.

This temporary binding is needed to limit credential sharing. At the local peer level, concurrent sessions using the same registered identity are treated as suspicious and communication is restricted to the first active session. This mechanism is not intended to provide a perfect solution to identity sharing, but it raises the difficulty of reusing the same registration proof across many active nodes.

\subsection{Message dissemination}
\label{sec:message-dissemination}

The protocol uses gossip-based dissemination to exchange governance and funding objects between peers. There is no central server that determines the system's global state. Instead, peers gradually spread issues, votes, solutions, campaigns, and pledges through the network. To reduce bandwidth usage, dissemination follows a push-pull mechanism based on \texttt{IHAVE} and \texttt{IWANT} messages.

Peers do not immediately send every full object to every neighbour. A peer first announces the IDs of objects it knows by sending an \texttt{IHAVE} message. The receiving peer compares these IDs with its local store and replies with an \texttt{IWANT} message for the objects it is missing. The original peer then sends only the requested objects. This avoids repeatedly transmitting full objects that a neighbour already has, which is important in a gossip network where the same object may arrive through multiple paths.

When a peer receives a protocol object, it validates the object before adding it to its local state. This validation checks the ID, signature, referenced objects, and object-specific rules. For example, a vote must be signed by a registered participant, must refer to an existing issue or solution, and must not duplicate an earlier vote from the same identity.

Some objects depend on parent objects. For example, a solution refers to an issue, a solution vote refers to a solution, and a funding pledge refers to a funding campaign. Because messages are disseminated asynchronously, a peer may receive a dependent object before it has received the object it depends on. Such objects are not immediately added to the main protocol state and are not advertised further. Instead, they are placed in an orphan pool indexed by the missing parent ID.

The orphan pool prevents peers from deriving inconsistent state from incomplete object chains. A solution cannot be evaluated before the corresponding issue is known, and a vote cannot be counted before the object being voted on has been validated. It also limits spam amplification: peers do not gossip dependent objects whose context is unknown, because forwarding them immediately would allow an attacker to spread large numbers of objects that cannot yet be fully verified.

When a missing parent object is later received and validated, the peer reconsiders the orphaned objects that depend on it. Each orphan is validated again in the now-complete context. Orphans should also be removed after a timeout or when the orphan pool reaches a size limit, preventing unbounded memory growth.

Because dissemination is asynchronous, the local state is inherently incomplete. A peer may learn about a solution before the corresponding issue, or about a campaign before all pledges have arrived. Missing local information is, therefore, not treated as evidence that an object does not exist. The protocol instead relies on explicit, validated objects and uses orphan pools until dependencies are available. Combined with immutable object IDs, this gives eventual convergence without requiring immediate global consistency. The \texttt{IHAVE}/\texttt{IWANT} exchange reduces redundant bandwidth usage, while orphan pools prevent incomplete dependencies from immediately affecting local state or being amplified through the network.

\subsection{Funding campaigns}
\label{sec:funding-campaigns}

Funding campaigns form the payment layer of the second protocol phase. They are designed to solve a specific coordination problem: once a developer submits a concrete solution, the community needs a way to compensate that work without handing funds directly to a group of maintainers, locking money in a long-lived contract, or making payment itself decide acceptance. The campaign is therefore linked to a submitted solution, but kept separate from the solution object. The solution remains independent, so funding attempts can expire, be retried, or be replaced with different payment terms.

\subsubsection{Design motivation}

A natural option would be to implement this mechanism as a smart contract. This design deliberately avoids that route because a smart contract would move the coordination logic onto a shared on-chain execution layer. Every pledge, deadline check, campaign update, and payout decision would become part of a globally replicated computation that must be ordered, executed, and stored by the underlying blockchain. This is an unsustainable scaling model for a system where most actions are ordinary coordination messages rather than final payments. TwoStepDemocracy, therefore, keeps the campaign state within the peer-to-peer protocol and uses Bitcoin only for final settlement.

This makes the funding campaign an application-level object whose rules are checked by peers, not an on-chain contract. The campaign defines the payment terms for one submitted solution: the developer's asking price, payout address, and deadline height. The asking price defines the funding target, while the deadline height defines when honest peers stop accepting new pledges for that campaign.

\subsubsection{Campaign creation}

A solution must have an explicit campaign before it can be accepted. This also applies to solutions for which the developer does not request payment. In that case, the campaign is treated as a free campaign, and no Bitcoin settlement is required. For paid campaigns, the solution is considered funded only once the requested amount has been settled in Bitcoin.

This explicit campaign requirement matters in a gossip-based system. The absence of a campaign in one peer's local state does not prove that no campaign exists; it may only mean that the campaign has not yet been received. By requiring solutions to refer to a concrete campaign object, peers can evaluate the solution, its requested compensation, and its funding status against an explicit protocol object rather than against missing or implicit state.

\subsubsection{Non-custodial pledges}

The funding mechanism is non-custodial. Contributors do not transfer money into an intermediate account controlled by the protocol or by another participant. Instead, they pledge existing Bitcoin outputs that they control. A pledge authorises the pledged output to be used in a future funding transaction for the campaign. Until that transaction is created and confirmed, the contributor remains in control of the output and can effectively revoke the pledge by spending it elsewhere.

To make pledge aggregation possible, pledges are signed using \texttt{SIGHASH\_ALL | SIGHASH\_ANYONECANPAY}. The \texttt{SIGHASH\_ALL} part fixes the transaction outputs, ensuring that the developer payout and campaign commitment cannot be changed after a contributor signs. The \texttt{ANYONECANPAY} part allows each contributor to sign only their own input, so that pledges from multiple contributors can later be combined into one final transaction. This allows a campaign to be funded collectively without requiring all contributors to coordinate at the same time.

\subsubsection{Pledge validity}

Pledge validity is dynamic. A pledge only counts toward the funding total while the pledged output remains unspent and the pledge remains compatible with the campaign terms. Since a contributor can revoke a pledge by spending the pledged UTXO elsewhere, funding totals are derived from currently valid pledges rather than stored as a fixed state.

The deadline height is also enforced at the application level. Honest peers stop accepting or counting pledges after the deadline, but a signed pledge remains technically usable as long as its UTXO remains unspent. The protocol, therefore, treats the deadline as a rule for campaign validity inside TwoStepDemocracy, not as a Bitcoin-level spending restriction.

\subsubsection{Final settlement}

Once the valid pledges cover the asking price and the required transaction fee, any peer can combine them into the final funding transaction and broadcast it. The original contributors do not need to be online at that moment. The miner fee is the difference between the selected input sum and the fixed campaign outputs.

The final transaction contains a commitment to the campaign in the metadata. This commitment binds the Bitcoin payment to the application-level funding campaign. Peers can therefore verify that a confirmed Bitcoin transaction corresponds to the intended campaign rather than to an unrelated payment. In the current design, this commitment is computed in a similar way as the registration commitment:
\[
    H(\texttt{protocol version} \mid \texttt{network} \mid \texttt{campaign ID})
\]

The funding campaign completes only after Bitcoin settlement. However, settlement alone does not mean that the solution is accepted. The corresponding solution must still satisfy the solution vote described above. This preserves the central separation of the design: funding measures whether the requested compensation has been matched, while voting determines whether the submitted solution has community approval.

\subsection{Design trade-offs}
\label{design-limitations}

The design intentionally keeps the coordination mechanism narrow. This is not only a simplification of the prototype, but also a design choice. Earlier alternatives quickly introduced heavier mechanisms, such as pay-to-vote, expert juries, appeal procedures, reputation systems, or stronger identity assumptions. These mechanisms address real problems, but they also make the system harder to understand, implement, and evaluate.

TwoStepDemocracy therefore makes several deliberate trade-offs. Voting is treated as a signal of approval, not as proof of technical correctness. Funding measures willingness to support a solution, not its technical merit. Open solution submission allows developers to compete, but it also leaves copycat or duplicate solutions as a social problem rather than a protocol-level one. These boundaries keep the core workflow simple: users express demand, developers submit solutions, the community votes, and funding remains separate from approval.

The implications of these trade-offs, together with future extensions such as stronger identity and review mechanisms, are discussed in Section~\ref{chapter:limitations-and-future-work}.

\begin{figure*}[!b]
    \centering
    \includegraphics[width=0.95\textwidth]{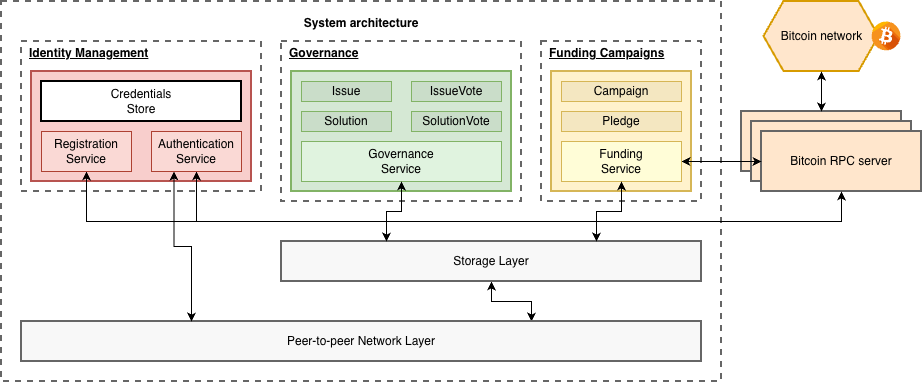}
    \caption{System architecture detailing the internal modules, services, and models as well as the external dependency on a Bitcoin RPC server.}
    \label{fig:system-architecture}
\end{figure*}

\section{Implementation}
\label{chapter:implementation}

The prototype is implemented in Python, targeting Python 3.12 while also being tested on Python 3.10. The peer-to-peer (P2P) messaging layer is built on top of IPv8\footnote{\href{https://py-ipv8.readthedocs.io}{py-ipv8.readthedocs.io}}, which provides the overlay networking functionality used for communication between nodes. Persistent local storage is implemented using SQLite\footnote{\href{https://sqlite.org}{sqlite.org}}, allowing each peer to store validated protocol objects and derive local state from them. The user interface is implemented in Qt\footnote{\href{https://www.qt.io}{qt.io}}, providing a desktop frontend for interacting with the protocol.

This section describes how the design from Chapter~\ref{chapter:design} is realised in the prototype. We first give an overview of the system architecture and protocol object model, after which we discuss identity, governance actions, funding campaigns, message dissemination, storage, and the frontend implementation in more detail.

\subsection{System architecture}

Figure \ref{fig:system-architecture} shows the main components of the prototype. Each instance of the application runs as a peer node. A peer node contains modules for identity, governance, funding, P2P networking, and storage. These modules operate on the same set of protocol objects: registrations, issues, solutions, votes, funding campaigns, and funding pledges.

The implementation follows the object-based design described in Section~\ref{sec:protocol-objects}. Protocol actions are represented as immutable objects that can be signed, validated, stored, and exchanged between peers. The local state of a node is derived from the valid objects it has received, rather than from a central database or shared mutable state. This allows each peer to independently reconstruct values such as vote counts, known solutions, and campaign funding progress.

The architecture separates protocol logic from external infrastructure. Governance objects are disseminated through the peer-to-peer layer, while Bitcoin-specific operations are performed through an external Bitcoin RPC node. This node can be run by the user or provided by another party. It is used to verify registration payments, inspect pledged UTXOs, and broadcast funding transactions, but it does not store or coordinate the governance state of the application.

\subsection{Bitcoin RPC server}
\label{sec:bitcoin-rpc-server}

The prototype interacts with the Bitcoin network through an external Bitcoin Core RPC node. This node can be run locally or provided by another party. The application uses the Bitcoin Network only to query data when the protocol requires external evidence, such as registration payments, pledged UTXOs, funding transactions, or confirmation status.

The RPC client is implemented as a small wrapper around Bitcoin Core's JSON-RPC interface. It handles endpoint configuration, authenticated requests, response validation, and error conversion. The client is wallet-agnostic for verification, so it can check transactions, block height, and UTXO state without requiring access to a user's wallet or private keys.

For development and testing, the prototype includes a wrapper script around \texttt{bitcoind} and \texttt{bitcoin-cli}. The script starts a local regtest node, creates or loads a demo wallet, mines blocks on demand, resets the local chain state, and provides convenience commands for demo transactions and PSBT signing. This makes the Bitcoin-dependent parts of the prototype reproducible without using mainnet coins or public infrastructure.

\begin{figure*}[!b]
    \centering
    \includegraphics[width=0.95\linewidth]{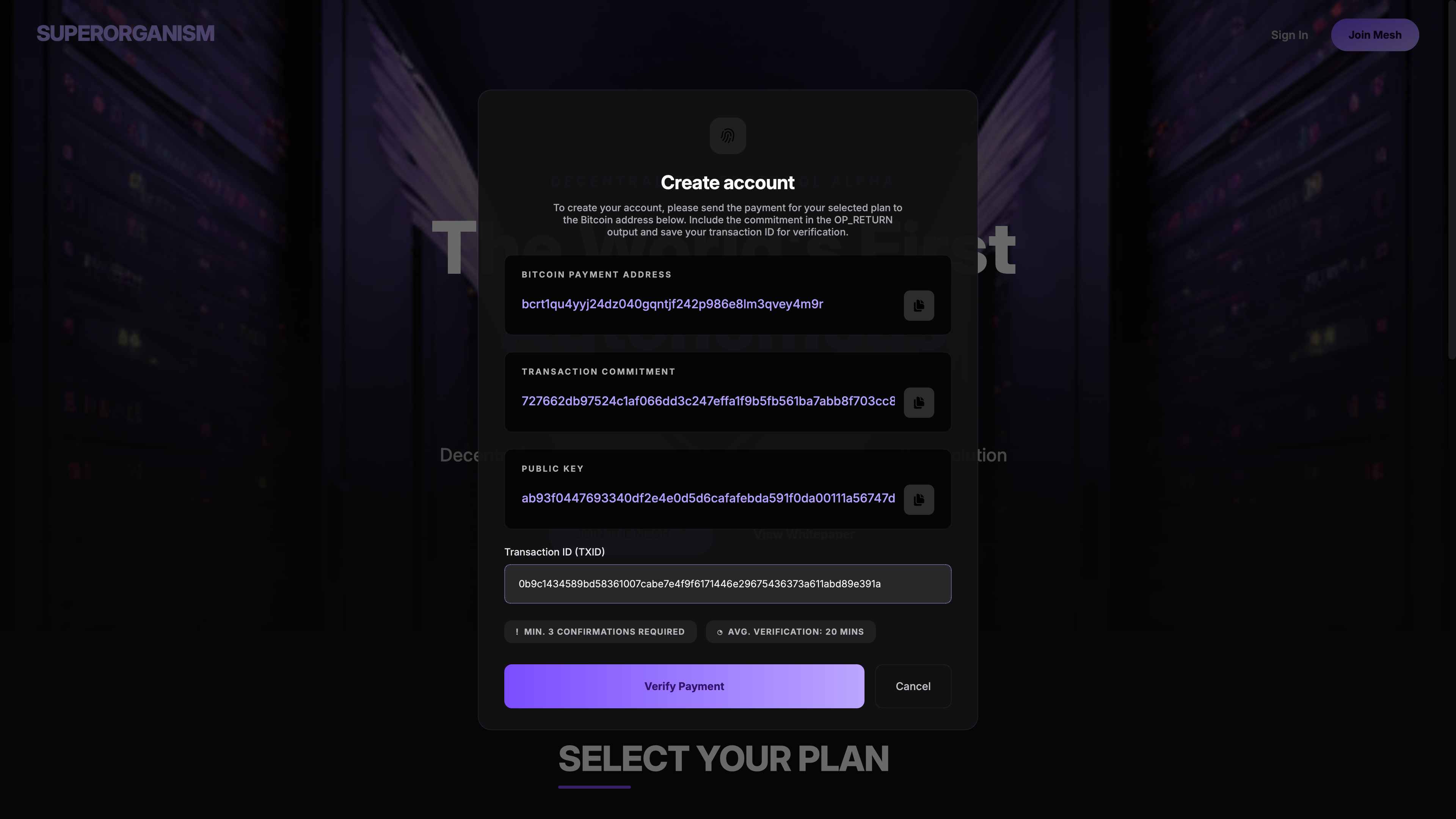}
    \caption{Registration flow in the prototype. The participant creates an application-level key pair and pays the 50,000 satoshi registration fee to the treasury address. The Bitcoin transaction includes an \texttt{OP\_RETURN} commitment binding the registration to the public key.}
    \label{fig:registration-flow}
\end{figure*}

\subsection{Identity}

The prototype implements the registration and login parts of the identity mechanism. The goal of this implementation was to validate the main design idea: a participant can be represented by an application-level public key, and this public key can be linked to a Bitcoin payment that serves as proof of registration.

\subsubsection{Registration}

The payment verification logic was implemented behind an interface, so that the underlying verification method can be changed later. This is useful because different options are possible. A production system could verify payments via community-operated Bitcoin nodes, seedboxes that perform network verification, or another mechanism agreed upon by the community. For the prototype, we used a local Bitcoin regtest environment, enabling us to test the full registration flow without relying on external services.

The regtest environment is managed using a Bash script around Bitcoin Core's \texttt{bitcoind} and \texttt{bitcoin-cli}. The script starts and stops the local node, creates test wallets, mines blocks, sends transactions, and exposes the treasury address used during registration. It also provides a reset command, which makes integration tests easier because each test can start from the same blockchain state.

During registration, the user is shown the treasury address and is prompted to make a payment to it. The transaction includes a metadata commitment of the form:
\[
    \texttt{sha256}(\texttt{VERSION} \mid \texttt{NETWORK} \mid \texttt{PUBLIC\_KEY})
\]

This commitment is stored in the transaction using the \texttt{OP\_RETURN} field. After the transaction is created, the application verifies that the transaction exists, pays to the expected treasury address, and contains the expected commitment. If these checks succeed, the application stores the account data needed for later authentication. Figure~\ref{fig:registration-flow} shows how it is all implemented in the prototype.

This implementation allowed us to test the registration mechanism end-to-end. However, the current prototype only demonstrates the concept in a local development environment. The final system would need a more robust deployment model for transaction verification, preferably one that does not require every participant to run a full Bitcoin node.

\subsubsection{Authentication}

\begin{figure*}[t]
    \centering
    \includegraphics[width=0.95\linewidth]{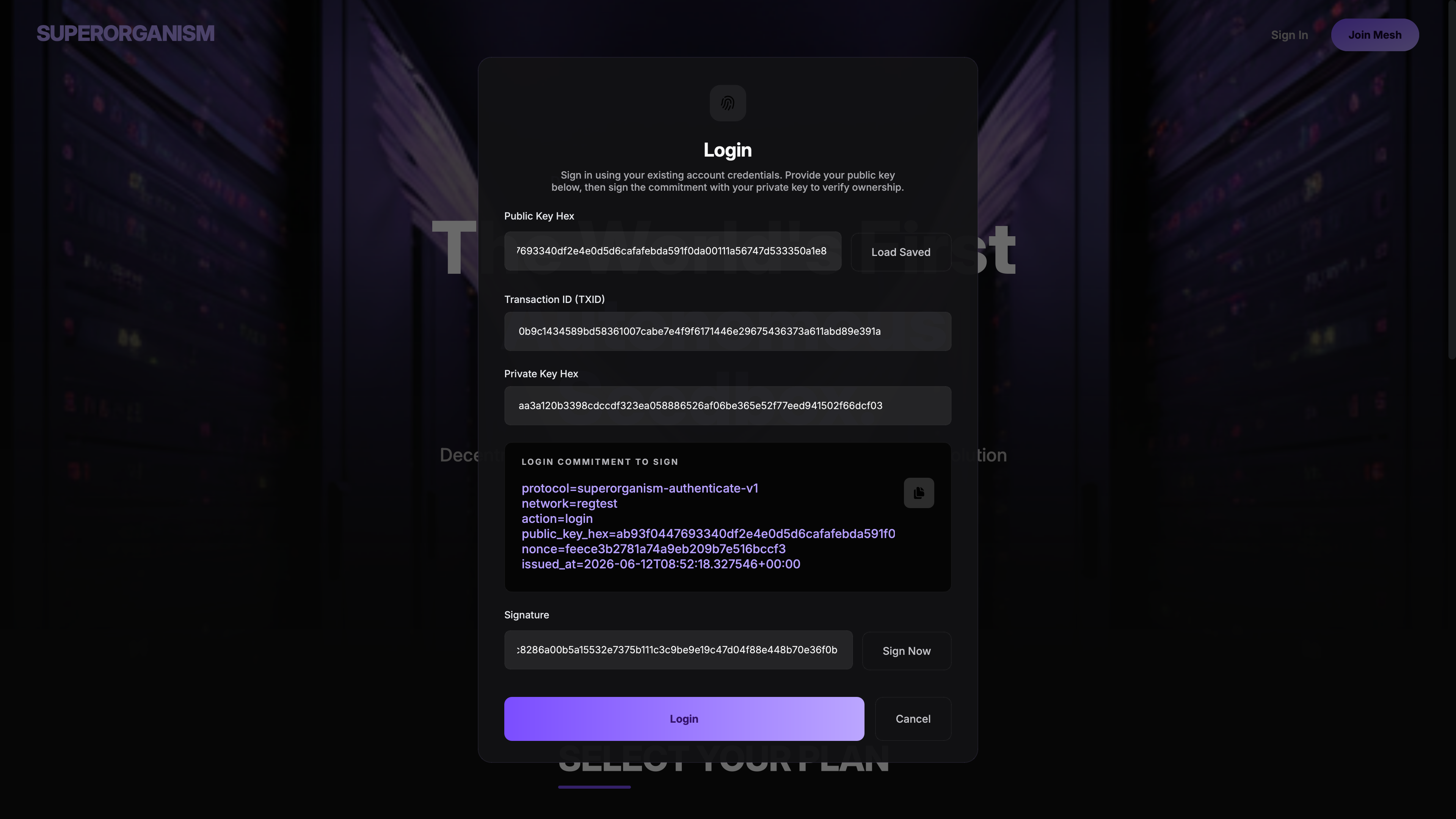}
    \caption{Authentication flow. The peer first verifies the registration transaction and then issues a fresh challenge. The user signs this challenge with the private key corresponding to the registered public key.}
    \label{fig:authentication-flow}
\end{figure*}

The login flow verifies two properties. First, it checks that the user provides a transaction identifier and public key that correspond to a valid registration. Second, it checks that the user controls the private key associated with that public key.

To prove key ownership, the application creates a fresh commitment containing the protocol name, network, action, account identifier, nonce, and issue time. The user signs this commitment with the private key belonging to the registered public key. The application then verifies the signature before accepting the login attempt. This prevents a user from logging in with only copied public registration data.

The prototype includes a simple user interface for demonstrating this flow, as can be seen in Figure~\ref{fig:authentication-flow}. The interface can load account data from a local file and use it to perform authentication. This was useful for quickly testing the concept, but it is not suitable for a secure production system. In the current prototype, the private key is stored in a plaintext JSON file alongside the rest of the account data. In a complete implementation, the private key should instead be stored in a hardware wallet, an encrypted key store, or another protected signing component. The application should request signatures from this component rather than handling the private key directly.

\subsubsection{Communication}

The prototype does not implement identity-aware peer communication. While the design includes a challenge-response flow for authenticating remote peers, this was left outside the scope because the networking layer is shared with the broader peer-to-peer application structure. Within the available time, the implementation focused on the complete registration and local authentication workflow. Communication-level authentication remains a clear extension point for future work.

\subsection{Issues, solutions, and votes}

Issues, solutions, issue votes, and solution votes are implemented as separate protocol objects. Each object is created locally, signed by the participant's application-level key, validated, stored, and then disseminated.

\begin{figure*}[t]
    \centering
    \includegraphics[width=0.95\linewidth]{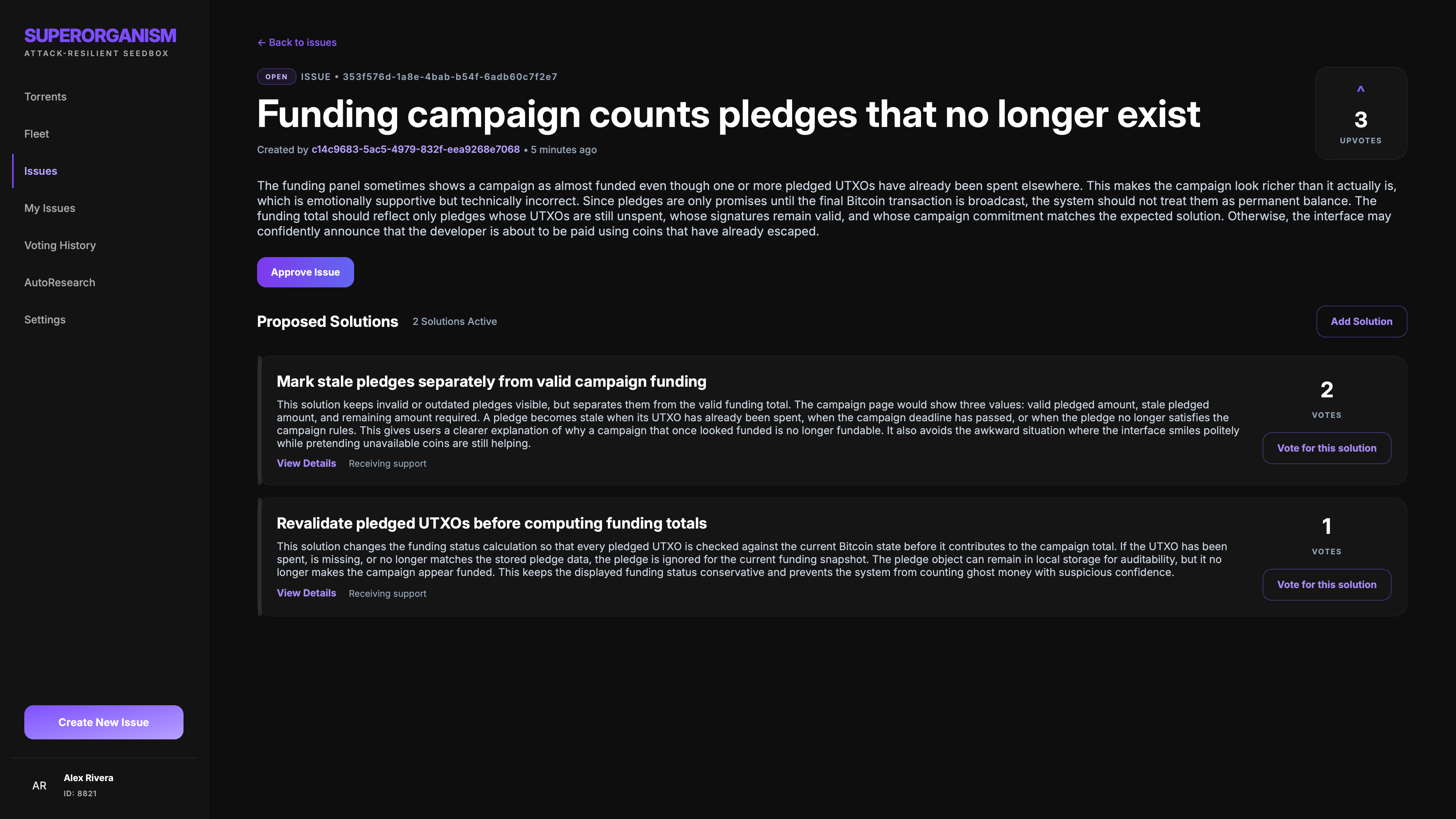}
    \caption{Issue detail view in the prototype frontend. The page shows an issue, its current support level, and the submitted solutions linked to it.}
    \label{fig:issue-page}
\end{figure*}

Although peer connections are authenticated, this is not sufficient for governance objects. A malicious peer could otherwise broadcast a vote while claiming that it came from another registered participant. For this reason, governance objects require object-level signatures. When a peer receives an object, it verifies the object's signature before accepting it. Objects with invalid signatures, unknown identities, or duplicate votes are rejected and are not forwarded further. Figure~\ref{fig:issue-page} shows some of these objects are displayed on an issue page.

\subsection{Funding campaigns}

Funding campaigns are implemented as a separate service layer. All its Bitcoin-specific checks are performed through the RPC client described in Section~\ref{sec:bitcoin-rpc-server}.

When a developer submits a solution, the app simultaneously creates a corresponding funding campaign. The campaign includes a Bitcoin payout address, an asking price, and a deadline height. If no payment fields are provided, the implementation creates a free campaign.

The pledge flow is split into request creation and signed pledge submission. To create a pledge request, the contributor provides a UTXO identified by \texttt{txid:vout}. The service checks that the referenced UTXO is unspent and sufficiently confirmed. It then creates a PSBT with the developer payout and the campaign commitment. This PSBT is returned to the user for external signing with \texttt{ALL|ANYONECANPAY}.

When the signed PSBT is submitted, the service validates it using the RPC client. If that succeeds, the pledge is stored and disseminated. Duplicate pledges for the same campaign and non-valid pledges are rejected by the repository.

Once enough valid pledges are available, any user can prepare the final transaction. They first validate and finalise the usable pledges, then select a subset whose total value covers the asking price and fee buffer. The selection heuristic prefers fewer inputs to reduce transaction size: if one pledge can cover the remaining shortfall, it chooses the smallest such pledge; otherwise, it chooses the largest remaining pledge. The selected signed inputs are then combined into a final raw transaction and sent to the Bitcoin network. The original pledgers do not need to be online at this point, since they already signed their individual pledge inputs.

\begin{figure*}[t]
    \centering
    \includegraphics[width=0.95\linewidth]{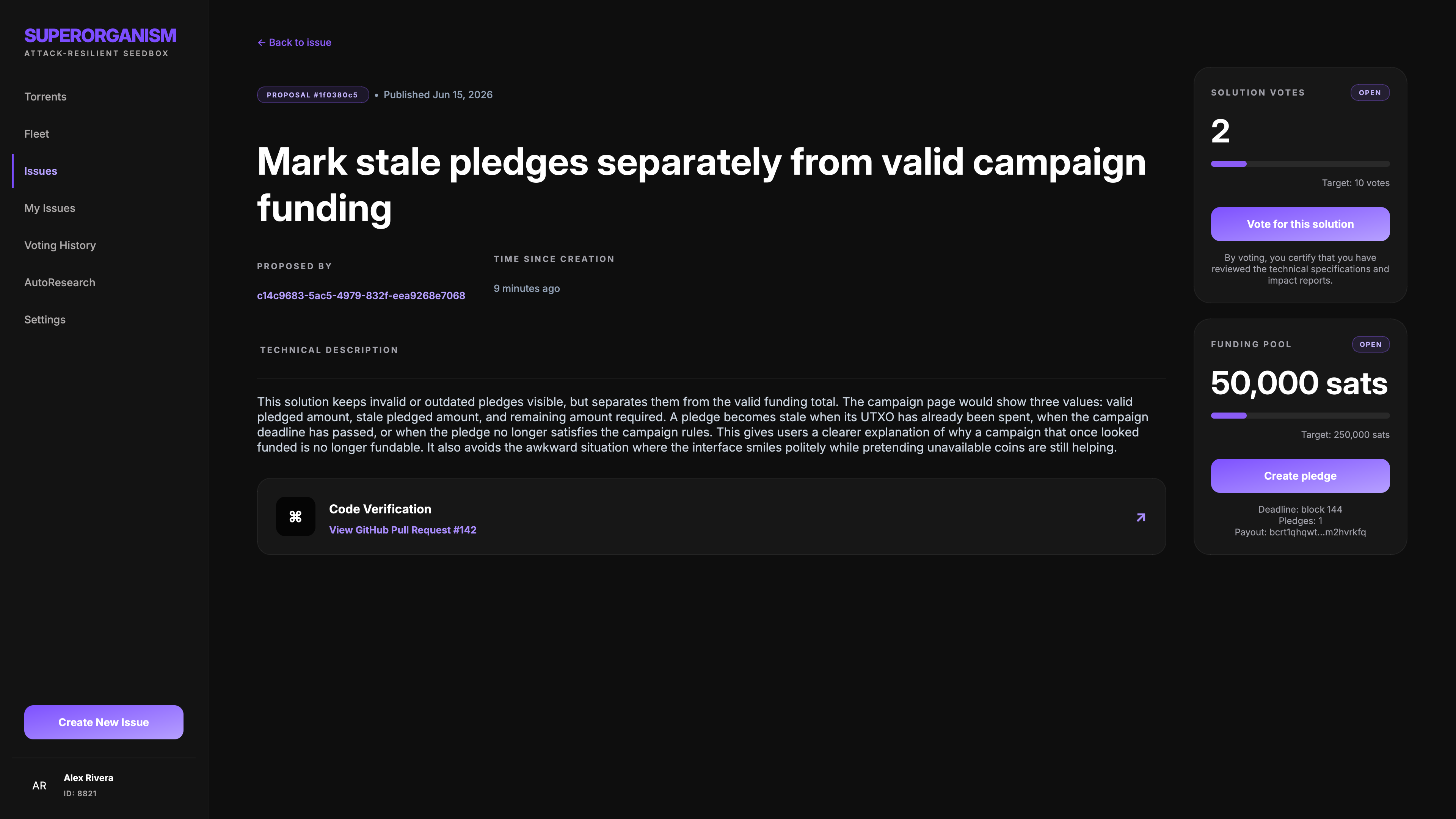}
    \caption{Solution detail view in the prototype frontend. The page shows a submitted solution together with its solution vote and linked funding campaign. The side panels make the two acceptance conditions visible: community approval through votes and economic support through pledged funding.}
    \label{fig:solution-page}
\end{figure*}

The current prototype recomputes funding status whenever the frontend refreshes. This makes the behaviour simple and conservative, because every displayed total is derived from the current Bitcoin state. However, it also means that old pledges are repeatedly revalidated. This is feasible for a prototype, but too expensive for a larger deployment, especially if Bitcoin RPC nodes become shared network resources. A production implementation should cache funding state and recompute defensively. Figure~\ref{fig:solution-page} shows the prototype UI of the solution page.

\subsection{Message dissemination}

Peer-to-peer communication is implemented using IPv8. The networking layer runs in a separate thread, so that message handling and periodic communication do not block the rest of the application. The dissemination layer is responsible for exchanging protocol objects between peers by forwarding knowledge about validated objects through \texttt{IHAVE} and \texttt{IWANT} messages.

The \texttt{IHAVE} exchange runs periodically. The interval is configurable; in the prototype it is set to one minute. At every interval, a peer advertises the IDs of known objects. The receiving peer compares these IDs with its local store and replies with an \texttt{IWANT} message for the objects it is missing. The sender then transmits the requested full objects one by one.

There is a special case when the object was just created locally, then direct peers cannot have received it yet. Therefore, newly created objects are sent immediately to the node's direct peers. They validate and store the object, but do not immediately forward the full object. Instead, the object ID is included in later \texttt{IHAVE} messages. This speeds up the dissemination of newly created objects.

Received objects are passed through the same validation pipeline as locally created objects. If an object is valid and its dependencies are known, it is stored and may be advertised in later gossip rounds. If a dependency is missing, the object is placed in the orphan pool until the parent object is received. Invalid objects are discarded and are not advertised further.

Object IDs in \texttt{IHAVE} messages are packed into bits to reduce overhead. Each advertised ID uses three bits to encode the protocol object type, followed by the object's 128-bit UUID4. One advertised object, therefore, requires $3 + 128 = 131$ bits. The maximum UDP payload size is conservatively set to 1300 bytes. This gives a maximum of

\begin{equation}
    \left\lfloor \frac{1300 \cdot 8}{131} \right\rfloor = 79
\end{equation}

object IDs per \texttt{IHAVE} message.

The current prototype includes all object types in the same \texttt{IHAVE} payload. If more IDs are available than can fit in a single message, the payload is split across multiple messages. This is sufficient for the prototype, but it is not yet an optimal dissemination policy. A future version could use separate \texttt{IHAVE} messages per object type and randomly select advertised IDs to obtain probabilistic dissemination guarantees, thereby reducing repeated transmission of the same local view.

\begin{figure}[h]
    \centering
    \includegraphics[width=\linewidth]{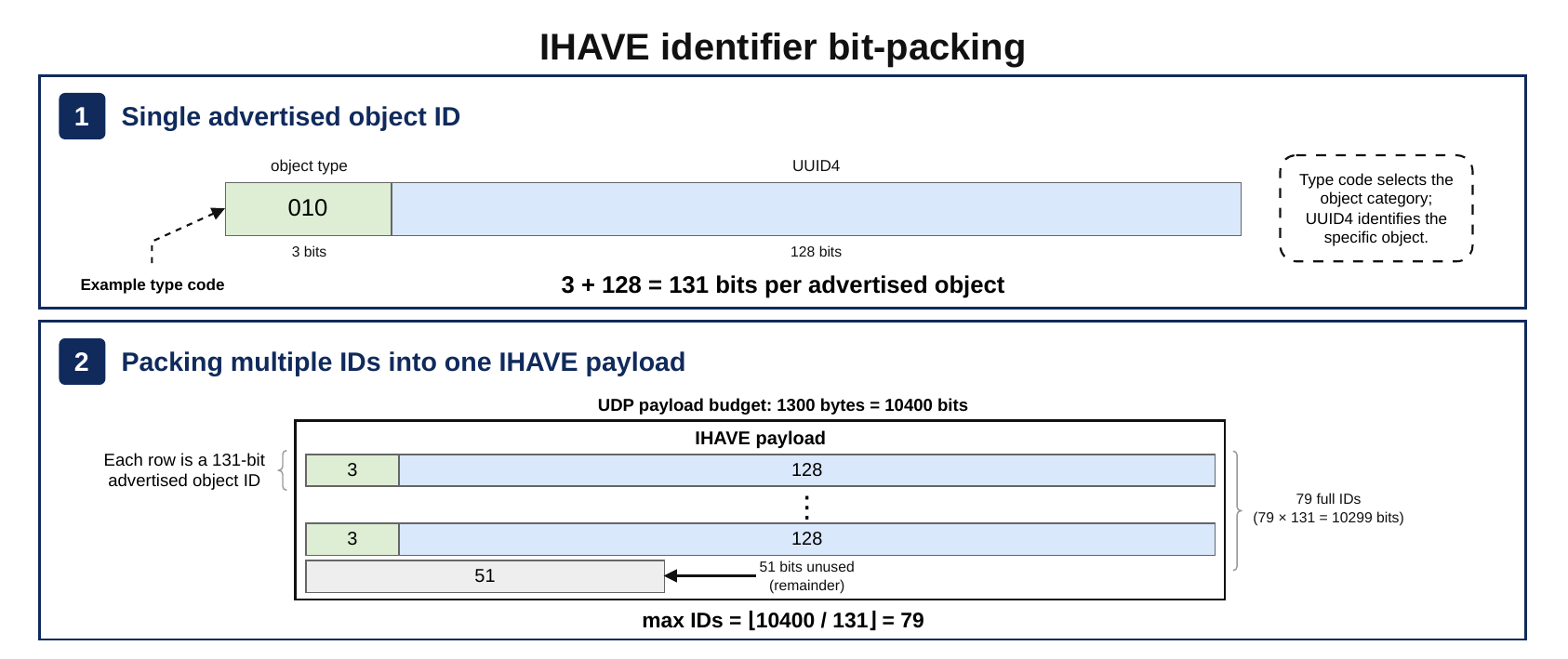}
    \caption{Bit-packing of object identifiers in an \texttt{IHAVE} message.}
    \label{fig:ihave-bit-packing}
\end{figure}

\subsection{Storage and derived state}

The prototype uses SQLite for local persistence through Python's built-in \texttt{sqlite3}\footnote{\href{https://docs.python.org/3/library/sqlite3.html}{https://docs.python.org/3/library/sqlite3.html}} module. Each peer stores the issues, votes, solutions, funding campaigns, and pledges it has received via the P2P dissemination layer and validated (using data from the Bitcoin network). The database represents the local view of a single peer instead of a shared global source of truth.

The implementation avoids storing derived values as authoritative state. For example, issue popularity is computed from issue vote objects. Similarly, campaign funding progress is derived from the currently valid pledges according to their Bitcoin transaction states. This keeps the storage layer close to the object-based protocol design.

The current prototype includes issue and solution descriptions directly in the IPv8 messages. To keep message sizes bounded, descriptions are limited in length. This is sufficient for the prototype, but it is restrictive for a real update proposal, where a solution may require long technical descriptions, build artefacts, or other files. The current implementation is built to be future-proof by already storing descriptions by content hash. This setup leaves a clear path toward IPFS-style\footnote{\href{https://ipfs.tech/}{https://ipfs.tech/}} sharing of larger descriptions and update files. Instead of sending the full data in every IPv8 message, a future version could disseminate only a content hash and retrieve the corresponding content from a separate file-sharing layer, potentially using tags. The current implementation does not yet include this file-sharing layer, but the database schema was designed so that such a change would not require extensive redesign.

Local SQLite storage is a practical choice for a prototype, but it is not the strongest consistency model for a P2P governance system. Since each peer stores only its own local view, peers may temporarily disagree about their state. A blockchain-based voting layer could provide stronger global ordering and finality for governance actions. This was not implemented because the focus of the prototype is the governance protocol and a Bitcoin-based funding mechanism rather than the design of a (voting) ledger. SQLite, therefore, serves as a lightweight persistence layer for evaluating the protocol flow, while stronger replicated storage is left for future work.

\subsection{Frontend prototype}

The frontend is implemented in Qt6 using the official Python binding PySide6\footnote{\href{https://pypi.org/project/PySide6/}{https://pypi.org/project/PySide6/}}. It provides a desktop interface to a local peer node. The frontend does not function as a central service. Instead, it calls the local application logic, which stores objects, communicates with peers through IPv8, and queries the configured Bitcoin RPC node.

The goal of the frontend is to demonstrate the end-to-end feasibility of the protocol, rather than to provide a production-ready user experience. It exposes the main protocol objects and actions to facilitate system testing and inspection during development. The prototype frontend can be especially useful in future research for evaluating social responses to the proposed governance protocol.

\section{Performance and scalability evaluation}
\label{chapter:performance}

This chapter evaluates the prototype. The goal is to identify where the current design scales and where further work is needed. We evaluate four aspects: identity verification, the storage required by a replicated local governance state, the dissemination of protocol objects between peers, and the validation and settlement cost of Bitcoin-based funding campaigns.

\subsection{Identity}
\label{sec:performance-identity}

The identity mechanism is evaluated mainly for correctness rather than performance. Registration evidence is stored externally in Bitcoin transactions, so the prototype does not introduce local storage overhead for identities. The relevant question is therefore whether the implementation correctly creates identities, binds public keys to registration commitments, verifies Bitcoin transaction evidence, and authenticates users through signed challenges.

We evaluate this through unit and integration tests. Unit tests cover deterministic functionality such as key generation, signing, signature verification, commitment construction, and request validation. Integration tests cover Bitcoin-backed registration verification. Table~\ref{tab:identity-module-test-coverage} reports the resulting line coverage. The tests clearly validate the core registration and authentication workflow implemented in the prototype.

\begin{table*}[t]
    \centering
    \caption{Test coverage of the identity module.}
    \label{tab:identity-module-test-coverage}
    \begin{tabular}{p{0.28\linewidth} | p{0.32\linewidth} | p{0.18\linewidth} | p{0.12\linewidth}}
        \textbf{Module} & \textbf{Main responsibility} & \textbf{Test strategy} & \textbf{Line coverage} \\
        \hline
        \texttt{crypto/} & Provides Ed25519 key generation, message signing, and signature verification primitives. & Unit tests & 100\% \\
        \hline
        \texttt{identity/} & Generates and represents application-level identities based on public/private key pairs. & Unit tests & 100\% \\
        \hline
        \texttt{models/} & Defines registration and authentication request models and validates their fields. & Integration tests & 100\% \\
        \hline
        \texttt{transaction\_verification/} & Verifies Bitcoin registration transactions, including payment evidence and transaction metadata. & Unit + integration tests & 99\% \\
        \hline
        \texttt{services/} & Coordinates registration and authentication workflows using commitments, signatures, stores, and transaction verification. & Integration tests & 84\% \\
        \hline
        Utilities & Provides constants, hex parsing, and registration commitment construction. & Unit tests & 100\% \\
    \end{tabular}
\end{table*}

\subsection{Storage requirements}
\label{sec:storage-requirements}

We evaluate the storage requirements of the prototype by measuring the size of the local SQLite database after inserting synthetic governance workloads. The goal of this experiment is to determine how much storage a peer needs under controlled usage scenarios. This is relevant because the current prototype follows a fully replicated storage model in which every peer eventually receives and stores all known governance objects.

For each workload, we create a fresh temporary SQLite database and insert the generated objects through the repository interface. After insertion, the database is checkpointed and compacted using SQLite's WAL checkpointing and \texttt{VACUUM}. We then compute the logical database size as \texttt{page\_size} multiplied by \texttt{page\_count}. This value represents the compacted SQLite storage required for the stored data and its associated tables and indexes.

We also measure an empty database separately. This captures the fixed storage overhead of the schema itself. For each workload, we report the net storage cost by subtracting this empty-database baseline from the logical database size. Since this fixed baseline is small and constant across workloads, the discussion focuses on the growth of the logical database size as the number of stored governance objects increases.

The workload is parameterised as a function of the number of users $N$:

\begin{equation}
    N \in \{1, 10, 100, 1000, 10000\}
\end{equation}

For each user level, the script generates a complete synthetic governance state according to three usage scenarios (light, moderate, or high participation). Each scenario defines the number of issues, solutions, votes, campaigns, and pledges relative to the user count. This allows us to compare different levels of activity without claiming that the measured results predict behaviour outside the evaluated range.

The scenarios used in the experiment are shown in Table~\ref{tab:storage_workload_scenarios}. The first rows define how many governance objects are generated relative to the number of users. For example, the issue fraction determines the number of generated issues, while the vote and pledge fractions determine how many users interact with each issue, solution, or campaign. The final rows define the synthetic payload sizes used for descriptions and signed pledge data. These payload sizes are kept fixed across scenarios so that the scenarios differ only in activity level, not in the assumed size of individual textual or funding objects.

\begin{table}[h]
    \centering
    \caption{Workload scenarios used in the SQLite storage experiment. Fractions are applied to the number of users $N$ and rounded up.}
    \label{tab:storage_workload_scenarios}
    \begin{tabular}{lccc}
        \toprule
        \textbf{Parameter} & \textbf{Light} & \textbf{Moderate} & \textbf{High} \\
        \midrule
        Issue fraction            & 0.01   & 0.05   & 0.10 \\
        Solutions per issue       & 2      & 2      & 2 \\
        Issue vote fraction       & 0.05   & 0.15   & 0.30 \\
        Solution vote fraction    & 0.03   & 0.10   & 0.20 \\
        Campaigns per solution    & 1      & 1      & 1 \\
        Pledge fraction           & 0.01   & 0.03   & 0.05 \\
        \midrule
        Issue description size    & 1000 B & 1000 B & 1000 B \\
        Solution description size & 2000 B & 2000 B & 2000 B \\
        Signed pledge PSBT size   & 1500 B & 1500 B & 1500 B \\
        \midrule
        Measured user levels $N$ & 
        \multicolumn{3}{c}{$\{1, 10, 100, 1000, 10000\}$} \\
        \bottomrule
    \end{tabular}
\end{table}

The exact values in Table~\ref{tab:storage_workload_scenarios} should not be interpreted as predictions of real-world use. In practice, it is difficult to determine such values in advance, because participation in TwoStepDemocracy depends strongly on the underlying application. If the application itself is active, then more users are likely to create issues, vote on proposals, and participate in funding. Conversely, if the underlying system has low adoption or little protocol evolution, the governance layer will also see less activity.

For this reason, we evaluate three scenarios instead of a single expected workload. The purpose of the experiment is therefore not to predict exact storage requirements for a future deployment, but to show how the storage footprint behaves under clearly stated and reproducible activity assumptions.

\begin{figure}[h]
    \centering
    \includegraphics[width=0.95\linewidth]{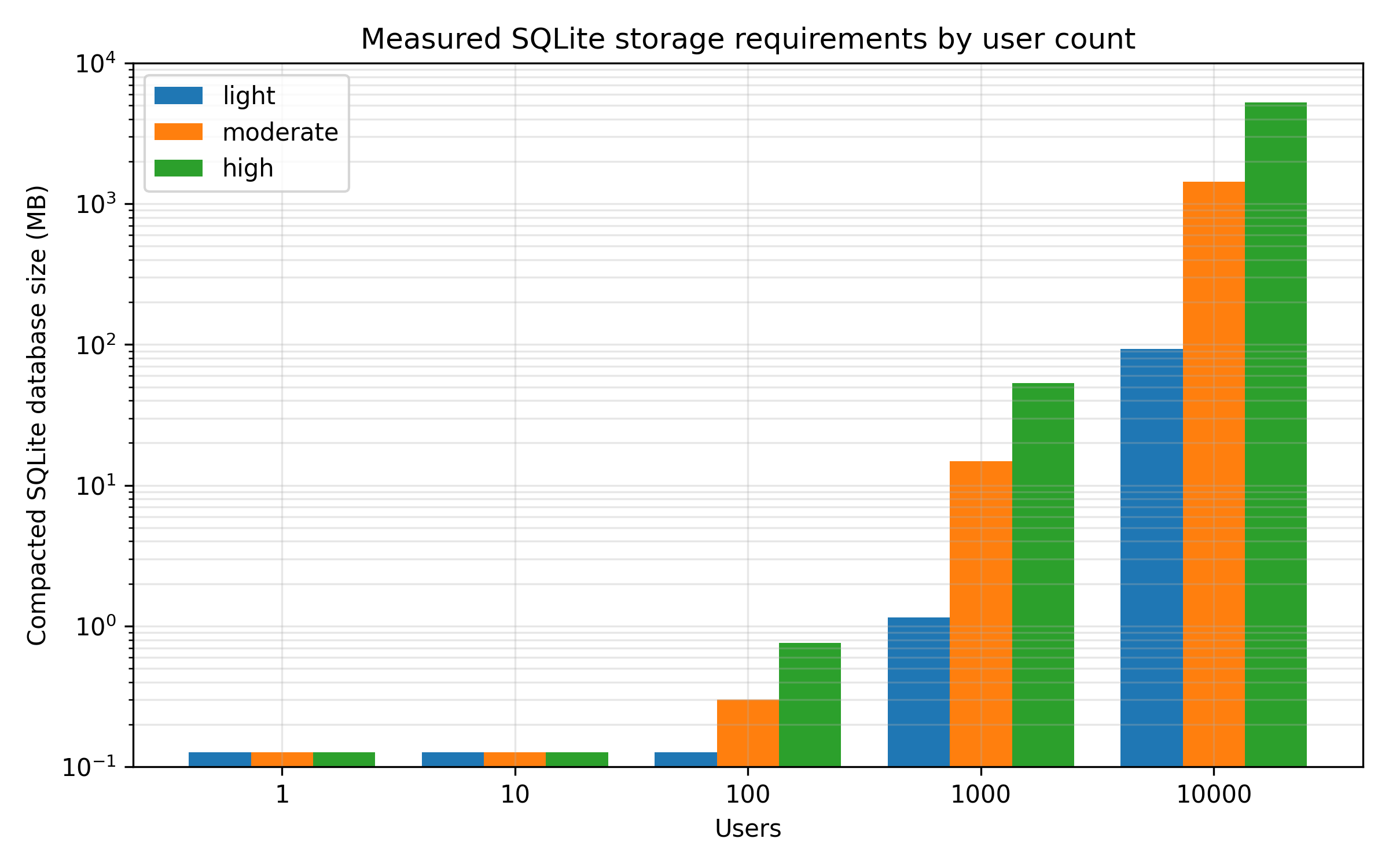}
    \caption{Measured SQLite storage requirements for fixed user levels and usage scenarios. Both axes use a logarithmic scale.}
    \label{fig:storage_growth}
\end{figure}

The result can be seen in Figure~\ref{fig:storage_growth}. The plot shows the measured logical database size against the number of users for each usage scenario. Both axes use a logarithmic scale because the evaluated user levels span several orders of magnitude.

The results show that the storage requirements remain small for low and moderate user counts, but grow quickly once the number of users and interactions increases. Under the light scenario, the database remains below 100 MB even at 10,000 users. Under the moderate scenario, however, the same user level already requires approximately 1.43 GB of logical storage. Under the high-participation scenario, this increases to approximately 5.26 GB. This difference illustrates that the storage cost is driven primarily by the number of interaction objects, especially votes and funding pledges, rather than by the number of issues alone.

These results suggest that the current fully replicated storage model is appropriate for the prototype and for smaller communities, but it is not a sufficient long-term storage strategy for very large deployments. If the system were used by communities with hundreds of thousands or millions of users, requiring every peer to retain every historical issue, vote, campaign, and pledge would lead to substantial local storage requirements. This is especially relevant because a production-ready decentralised implementation would likely add further overhead for verifiability, for example by anchoring objects in a blockchain or another append-only commitment structure.

\subsection{Dissemination overhead}
\label{sec:dissemination-overhead}

\begin{table*}[!b]
    \centering
    \caption{Test coverage of the funding campaign module.}
    \label{tab:funding-module-test-coverage}
    \begin{tabular}{p{0.20\linewidth} | p{0.34\linewidth} | p{0.18\linewidth} | p{0.18\linewidth}}
        \textbf{File} & \textbf{Main responsibility} & \textbf{Test strategy} & \textbf{Line coverage} \\
        \hline
        \texttt{bitcoin\_tx.py} & Parses, serializes, identifies, and combines Bitcoin transactions for \texttt{ALL|ANYONECANPAY} pledge aggregation. & Unit tests & 100\% \\
        \hline
        \texttt{models.py} & Defines immutable funding campaigns and pledges, validates object fields, and computes campaign commitments. & Unit tests & 100\% \\
        \hline
        \texttt{service.py} & Coordinates repository state, Bitcoin RPC calls, pledge validation, funding status, and final transaction construction. & Unit + integration tests & 89\% \\
    \end{tabular}
\end{table*}

We did not include a full empirical dissemination-latency evaluation, because the available hardware did not allow experiments at a network size large enough to produce meaningful scalability results. Instead, the goal is to show why the \texttt{IHAVE}/\texttt{IWANT} design is more bandwidth-efficient than naively pushing every full object to every neighbour.

Let \(S\) be the average size of a full governance object in bytes. We measured \(S = 925\) bytes on average assuming equal distribution of the different objects. Let \(I\) be the size of one advertised object identifier. In the prototype, each advertised identifier consists of a 3-bit object type and a 128-bit UUID, so

\begin{equation}
    I = \frac{3 + 128}{8} = 16.375 \text{ bytes}.
\end{equation}

With a conservative UDP payload budget of 1300 bytes, one \texttt{IHAVE} message can therefore contain

\begin{equation}
    \left\lfloor \frac{1300 \cdot 8}{131} \right\rfloor = 79
\end{equation}

object identifiers. This means that one control message can advertise 79 known objects without sending their full contents.

Consider a network of \(n\) peers with average degree \(d\). In a naive push model, a peer forwards every full object to all neighbours. Once the object has spread through the network, many peers receive the same object multiple times through different neighbours. Ignoring protocol headers, the total full-object traffic per object is approximately

\begin{equation}
    C_{\text{push}} \approx n d S.
\end{equation}

Only \(n-1\) of these transmissions are necessary to deliver the object to all peers for the first time. The remaining transmissions are duplicates caused by network connectivity. Each peer receives the object from roughly \(d\) neighbours, even though one successful delivery would have been enough.

In the push-pull model, full objects are only sent when a peer explicitly requests them with an \texttt{IWANT}. The full-object traffic is therefore closer to one successful delivery per receiving peer:

\begin{equation}
    C_{\text{full,push-pull}} \approx (n-1)S.
\end{equation}

The cost is that peers must exchange identifier advertisements. If an object identifier is advertised for \(r\) gossip rounds before all relevant peers have discovered it, the identifier traffic is approximately

\begin{equation}
    C_{\text{id,push-pull}} \approx r n d I.
\end{equation}

The total push-pull cost per object is therefore

\begin{equation}
    C_{\text{push-pull}} \approx (n-1)S + r n d I.
\end{equation}

The push-pull method is more efficient when

\begin{equation}
    (n-1)S + r n d I < n d S.
\end{equation}

For large \(n\), this simplifies to

\begin{equation}
    \frac{C_{\text{push-pull}}}{C_{\text{push}}}
    \approx
    \frac{1}{d} + r\frac{I}{S}.
\end{equation}

Using the measured estimate \(S = 925\) bytes and \(I = 16.375\) bytes, the identifier is only about \(1.8\%\) of the size of a full object. For example, with an average degree of \(d = 8\) and \(r = 10\) gossip rounds, the relative cost becomes

\begin{equation}
    \frac{C_{\text{push-pull}}}{C_{\text{push}}}
    \approx
    \frac{1}{8} + 10 \cdot \frac{16.375}{925}
    \approx 0.302.
\end{equation}

Under these assumptions, push-pull dissemination uses roughly \(30.2\%\) of the bandwidth of naive full-object pushing, before accounting for additional protocol headers. The exact value depends on object size, network degree, and the number of gossip rounds, but the trend is clear: replacing duplicate full-object transmissions with compact identifiers can substantially reduce bandwidth usage.

This analysis is not perfect and does not replace an empirical latency evaluation. It does, however, explain the main scalability advantage of the implemented dissemination strategy. The protocol trades repeated transmission of large objects for repeated transmission of compact identifiers. Since an identifier is much smaller than a full governance object, the push-pull approach is expected to reduce bandwidth substantially, especially in well-connected networks where naive pushing would create many duplicate full-object deliveries.

\subsection{Funding campaigns}
\label{sec:performance-funding-campaigns}

The funding campaign module is one of the most critical parts of the prototype, because it connects protocol-level governance objects to Bitcoin transactions. This part of the evaluation does not naturally produce many scalability graphs. Instead, the goal was to validate that the funding logic behaves correctly and show the costs associated with settling the final transactions.

\subsubsection{Validation}
The module consists of three files. The first two files, \texttt{bitcoin\_tx.py} and \texttt{models.py}, are deterministic and independent of external services. They were therefore tested with unit tests only, reaching 100\% line coverage. The third file, \texttt{service.py}, is more difficult to test because it depends on both the local repository and a Bitcoin RPC client. Without mocking either dependency, unit tests reached 42\% line coverage. This mostly covers local validation logic and branches that can be exercised without live repository and chain state.

To test the service more realistically, we added integration tests using a temporary SQLite database and a local Bitcoin regtest RPC server. With these integration tests, line coverage for \texttt{service.py} increased to 89\%. The remaining uncovered lines are mostly exceptional branches and bad-weather cases that depend on rare repository failures or Bitcoin RPC failures. Table~\ref{tab:funding-module-test-coverage} summarises the resulting coverage.

This result is important because the service layer contains the behaviour that cannot be validated by pure unit tests alone. Testing this against a regtest Bitcoin node gives stronger evidence than mocked RPC responses would provide, because the tests exercise the same class of chain-dependent behaviour that the prototype relies on during normal operation.

\subsubsection{Costs}
\label{sec:evaluation-many-small-pledges}

The funding mechanism allows a campaign to be funded by many independent pledges. This is useful for collective funding, but it also creates a practical cost: every pledge that is included in the final funding transaction contributes one Bitcoin transaction input. If a campaign is funded by many small pledges, the final transaction therefore becomes larger, which increases the miner fee required for settlement.

To evaluate this effect, we construct final funding transactions with an increasing number of pledged inputs. Each pledge contributes one input, while the transaction outputs remain fixed. For each pledge count, we measure the virtual transaction size and estimate the required fee using fixed fee rates of 1, 5, 10, and 15 sat/vB. Since the experiment runs on regtest, these fee rates are scenario values rather than fees selected by a live Bitcoin mempool. However, they are chosen to reflect realistic orders of magnitude. Galaxy Research reports that Bitcoin fee pressure had collapsed in 2025, with a meaningful share of blocks having an average fee rate of 1 sat/vB or less~\cite{GalaxyBitcoinFees2025}. We therefore treat 1 sat/vB as the current low-fee baseline, while 5, 10, and 25 sat/vB are increasingly conservative congestion scenarios. The estimated fee is computed as:
\[
    \texttt{fee\_sats} = \texttt{virtual\_size\_vB} \cdot \texttt{fee\_rate\_sat/vB}.
\]

To make the results easier to interpret, we also convert fees to US dollars using a conservative Bitcoin price of 60,000 USD. This value is deliberately below most 2026 prices and close to the lowest BTC/USD levels observed in the first half of 2026, where historical data shows lows around 59,200 USD~\cite{InvestingBitcoinHistoricalData}. The conversion is therefore:
\[
    \texttt{fee\_usd}
    =
    \frac{\texttt{fee\_sats}}{10^8}
    \cdot 60000.
\]
This conversion is not intended as a price prediction. It is only used to give the satoshi-denominated fees a more intuitive monetary scale.

We evaluate campaigns with up to 150 pledges. This upper bound is not a Bitcoin protocol limit: Bitcoin transactions are constrained by transaction weight rather than by a fixed maximum number of inputs. The bound was chosen to represent a large but still plausible campaign in which many users contribute small amounts. It is also sufficiently below standard relay-size boundaries, so the experiment focuses on ordinary fee growth rather than edge-case transaction validity.

\begin{figure}[h]
    \centering
    \includegraphics[width=0.95\linewidth]{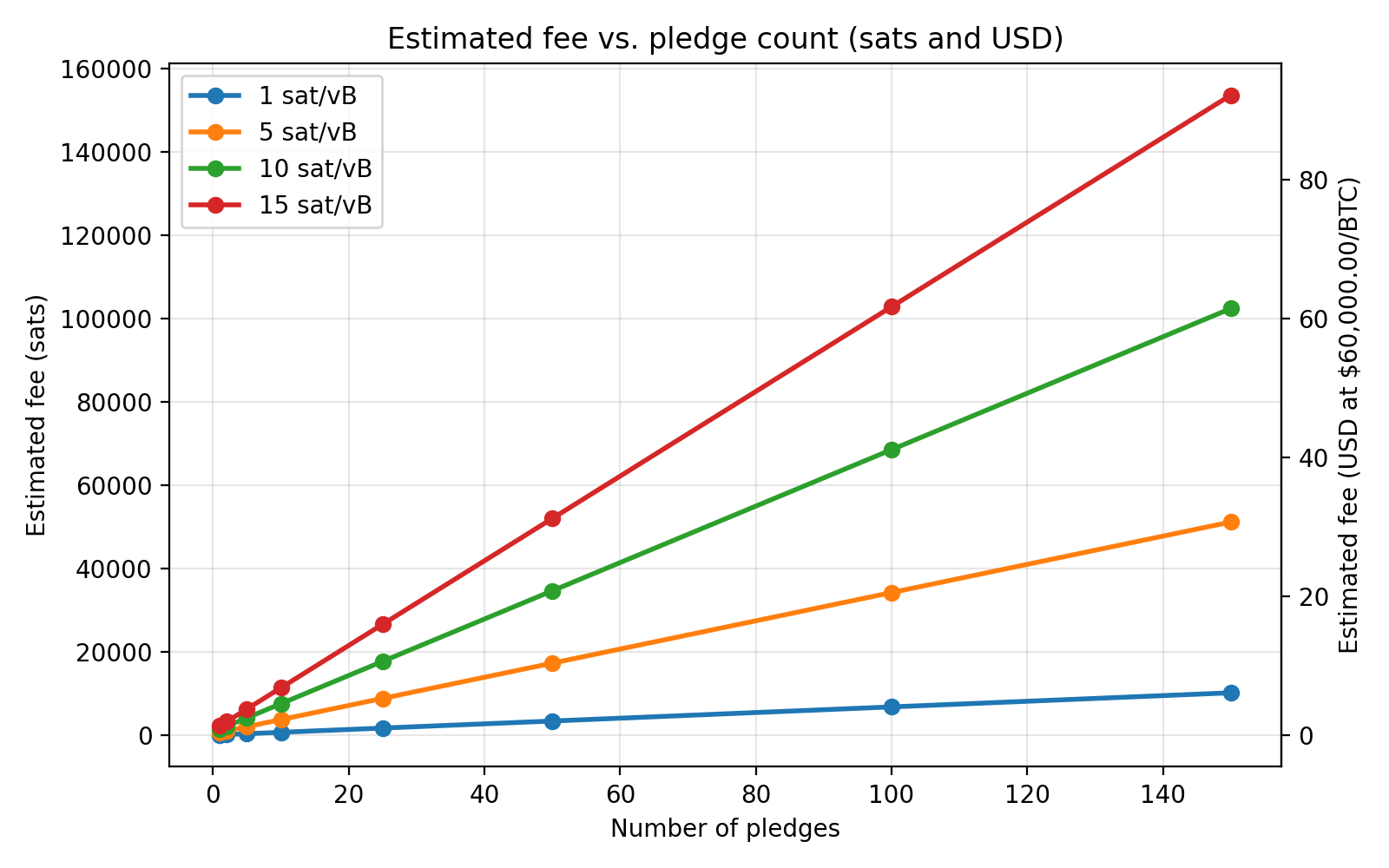}
    \caption{Estimated final transaction fee as the number of pledged inputs increases. Each pledge contributes one input to the final funding transaction, while the outputs remain fixed. The fee is estimated by multiplying the measured virtual transaction size by fixed illustrative fee rates of 1, 5, 10, and 15 sat/vB.}
    \label{fig:fee-vs-pledges}
\end{figure}

Figure~\ref{fig:fee-vs-pledges} shows the resulting fee estimates. The trend is linear: each additional pledge adds another input, which increases the virtual transaction size by approximately a fixed amount. As a result, the final fee grows proportionally with the number of pledges at any fixed fee rate. This confirms that many small pledges are technically possible, but not free. Campaigns funded by many small contributors may incur noticeably higher settlement costs than campaigns funded by fewer larger pledges.

This result does not undermine the funding design, but it clarifies one of its trade-offs. The protocol keeps pledges non-custodial and independently signed, which allows users to contribute without locking funds in a contract or coordinating online at the same time. The cost of that flexibility is that the final settlement transaction scales with the number of included pledge inputs. In practice, the frontend should therefore make estimated settlement fees visible, and peers preparing the final transaction should prefer pledge selections that cover the target with fewer inputs when possible.

\section{Limitations and future work}
\label{chapter:limitations-and-future-work}

We propose TwoStepDemocracy as a minimal coordination layer for decentralised software evolution. The evaluation shows that the core workflow can be implemented: governance actions can be represented as signed protocol objects, stored locally, exchanged between peers, and linked to non-custodial Bitcoin funding campaigns. These results support the technical feasibility of the design, but they do not yet establish that the system would work as a social governance mechanism. The remaining limitations and future work, therefore, fall into two categories: improving the technical prototype and validating whether real communities would use such a system effectively.

\subsection{Social validation}

The most important limitation of this thesis is that TwoStepDemocracy was evaluated primarily as a technical system. This was necessary within the scope of the thesis, but it leaves open the question of social viability. Governance also depends on how people behave when they are asked to vote, fund, review, compete, and trust each other. A meaningful evaluation would therefore require a social experiment with a sufficiently large user group.

Such a social experiment can also help determine the voting threshold. We did not try to determine the correct threshold for accepting solutions, because this is mainly a social design choice rather than a technical property. The threshold should likely depend on the community's size and activity. The prototype uses a fixed threshold to demonstrate the workflow. Future work should evaluate different threshold rules.

\subsection{Identity}

The current identity mechanism requires a registration fee, making identity creation costly. This raises the cost of Sybil attacks, but it does not prove that every account corresponds to a unique person. A wealthy attacker can still create multiple accounts, and determined users may still share account credentials. Future work should investigate stronger identity mechanisms.

The registration payment also raises a design question that we did not fully address: what should happen to the collected funds? The mechanism was originally designed in the context of a self-replicating seedbox system, where registration payments could naturally flow into a treasury used to maintain or expand the underlying infrastructure. In TwoStepDemocracy, however, this treasury function is not implemented. The payment is used only as evidence of costly registration. Future work could investigate whether these funds could be used towards the community. For example, registration revenue could subsidise storage and dissemination infrastructure, cover Bitcoin settlement fees, support public-good development, or be redistributed through future funding campaigns. This would turn the identity mechanism from a pure anti-Sybil cost into a source of shared protocol resources.

\subsection{Solution correctness and review}

A second limitation is that solution voting does not prove technical correctness. The prototype allows users to approve a submitted solution, but it does not verify that the solution is secure, maintainable, or that it actually implements the requested change. Future versions should therefore explore review mechanisms that are separate from ordinary solution voting.

A related limitation is duplicate or copycat solutions. Since solutions are submitted as open protocol objects, another developer may submit a very similar implementation with a lower asking price. The protocol can bind votes and funding to a specific solution identifier, but it cannot determine whether two submitted solutions are technically or intellectually equivalent. Handling this would require additional review, authorship, or dispute-resolution mechanisms.

\subsection{Storage and dissemination}

The current prototype assumes that peers eventually store all validated governance objects. This is simple and useful for experimentation, but it is not ideal for very large communities. The storage experiment showed that local storage grows with the number of users and interactions. A production system should therefore avoid requiring every peer to store the complete governance history forever \cite{CongRenPouwelse2019CHECO}. Future work should investigate archival roles, pruning, snapshots, and summarised historical state. The storage layer has been designed to easily support content-addressed storage, potentially using tags \cite{FokkerPouwelseBuntine2006P2PWikipedia}.

Future work should also investigate whether voting records should be anchored in a blockchain, similar to the public bulletin-board role used in SmartphoneDemocracy~\cite{JozwikPouwelse2025SmartphoneDemocracy}. In the current prototype, votes are stored only in local peer databases, so peers may temporarily disagree about which votes they have seen. A blockchain could serve as a shared, append-only record for the most important voting events, such as issue votes, solution votes, and final acceptance outcomes. This would improve ordering, auditability, and long-term verifiability.

\subsection{Toward full self-evolution}

Finally, TwoStepDemocracy only implements the coordination layer of a larger vision. It coordinates demand, approval, and funding, but it does not implement automatic distribution, deployment, rollback, or package management. These mechanisms are necessary for a system to truly evolve. Future work should therefore connect the coordination layer we developed to a broader self-updating infrastructure.

\section{Conclusion}
\label{chapter:conclusion}

This thesis started from a paradox. Decentralised systems are often built to resist central control, yet their evolution still depends on centralised platforms, fragile maintainer structures, and socially opaque decision processes. The code may run in a peer-to-peer network, but the power to decide what the code becomes often remains elsewhere.

The research question asked how a decentralised protocol can support its own evolution through a hybrid mechanism that combines democratic decision-making with public-goods funding, without relying on central maintainers to coordinate issues, approve solutions, and reward contributors. We answered that question by designing and implementing TwoStepDemocracy: a prototype coordination layer for decentralised software evolution.

The thesis contribution is the composition of several mechanisms into a single workflow. Costly registration provides a lightweight barrier against identity multiplication. Signed protocol objects make governance actions independently verifiable. Gossip-based dissemination allows issues, votes, solutions, campaigns, and pledges to spread without a central server. Solution voting separates approval from issue demand. Funding campaigns, the real pinnacle, separate economic support from voting power, while still giving developers a concrete path to compensation.

The implementation and evaluation show that this coordination layer is technically feasible. The prototype demonstrates the full path from issue creation to solution voting and Bitcoin-based funding. At the same time, the thesis should be understood as a technical proof of concept, not as proof that the governance model will work socially. Whether users vote carefully, whether developers trust the funding process, whether public-goods funding is sufficient, and whether the system produces legitimate decisions can only be answered through a larger social experiment. That experiment was outside the feasible scope of this thesis, but it is the natural next step.

Ultimately, the dream is that decentralised systems that no longer depend on external software forges to decide how they evolve. TwoStepDemocracy does not complete that vision. It does not remove human judgment, nor does it solve every problem of decentralised governance. What it does show is that protocol advancement can itself become part of the system. That is the first step toward decentralised software that does not simply run without a centre, but can also evolve without returning to one.

%% Add bibliography
\printbibliography[heading=bibintoc,title=References]

@article{AmritHillegersberg2010Coreperiphery,
  author       = {Chintan Amrit and Jos van Hillegersberg},
  title        = {Exploring the Impact of Socio-Technical Core-Periphery Structures in Open Source Software Development},
  journal      = {Journal of Information Technology},
  volume       = {25},
  number       = {2},
  pages        = {216--229},
  year         = {2010},
  doi          = {10.1057/jit.2010.7}
}

@article{ArrunadaGaricano2018Blockchain,
  author       = {Benito Arru{\~n}ada and Luis Garicano},
  title        = {Blockchain: The Birth of Decentralized Governance},
  journal      = {Pompeu Fabra University Economics Working Paper},
  number       = {1608},
  year         = {2018}
}

@misc{AvelinoConstantinouTulioSerebrenik2019Abandonment,
  author       = {Guilherme Avelino and Eleni Constantinou and Marco Tulio Valente and Alexander Serebrenik},
  booktitle    = { 2019 ACM/IEEE International Symposium on Empirical Software Engineering and Measurement (ESEM) },
  title        = {On the Abandonment and Survival of Open Source Projects: An Empirical Investigation},
  pages        = {1-12},
  year         = {2019},
  eprint       = {1906.08058},
  primaryClass = {cs.SE},
  doi          = {10.1109/ESEM.2019.8870181},
  url          = {https://doi.ieeecomputersociety.org/10.1109/ESEM.2019.8870181},
  publisher    = {IEEE Computer Society},
  address      = {Los Alamitos, CA, USA},
}

@article{BagnoliLipman1989ProvisionPublicGoods,
  author  = {Bagnoli, Mark and Lipman, Barton L.},
  title   = {{Provision of Public Goods: Fully Implementing the Core through Private Contributions}},
  journal = {The Review of Economic Studies},
  volume  = {56},
  number  = {4},
  pages   = {583--601},
  year    = {1989},
  doi     = {10.2307/2297502}
}

@misc{BIP2017PSBT,
  title        = {Partially Signed Bitcoin Transaction Format},
  author       = {Ava Chow},
  year         = {2017},
  url          = {https://bips.dev/174/},
  note         = {Accessed: 2026-05-25}
}

@misc{BIP3,
  author       = {Mark Erhardt},
  title        = {{BIP 3: Updated BIP Process}},
  year         = {2025},
  url          = {https://bips.dev/3/}
}

@misc{BitcoinDeveloper2025Transactions,
  title        = {Transactions},
  author       = {{Bitcoin Developer Documentation}},
  year         = {2025},
  url          = {https://developer.bitcoin.org/devguide/transactions.html},
  note         = {Accessed: 2026-05-25}
}

@inproceedings{BorgeKokorisJovanovicGasserGaillyFord2017Pop,
  author       = {Maria Borge and Eleftherios Kokoris-Kogias and Philipp Jovanovic and Linus Gasser and Nicolas Gailly and Bryan Ford},
  title        = {Proof-of-Personhood: Redemocratizing Permissionless Cryptocurrencies},
  booktitle    = {2017 IEEE European Symposium on Security and Privacy Workshops (EuroS\&PW)},
  year         = {2017},
  pages        = {23--26},
  publisher    = {IEEE},
  doi          = {10.1109/EuroSPW.2017.46}
}

@article{ButerinHitzigWeyl2019LiberalRadicalism,
  author       = {Vitalik Buterin and Zo{\"e} Hitzig and E. Glen Weyl},
  title        = {{A Flexible Design for Funding Public Goods}},
  journal      = {Management Science},
  volume       = {65},
  number       = {11},
  pages        = {5171--5187},
  year         = {2019},
  doi          = {10.1287/mnsc.2019.3337},
  url          = {https://doi.org/10.1287/mnsc.2019.3337}
}

@article{ChristoffGrossi2017LiquidDemocracy,
  author        = {Zo{\'e} Christoff and Davide Grossi},
  title         = {{Binary Voting with Delegable Proxy: An Analysis of Liquid Democracy}},
  journal       = {arXiv preprint arXiv:1707.08741},
  year          = {2017},
  eprint        = {1707.08741},
  archivePrefix = {arXiv},
  primaryClass  = {cs.GT},
  url           = {https://arxiv.org/abs/1707.08741}
}

@inproceedings{CongRenPouwelse2019CHECO,
  author    = {Cong, Kelong and Ren, Zhijie and Pouwelse, Johan},
  title     = {{A Blockchain Consensus Protocol with Horizontal Scalability}},
  booktitle = {{2018 IFIP Networking Conference (IFIP Networking) and Workshops}},
  editor    = {Stiller, Burkhard},
  pages     = {1-9},
  year      = {2019},
  publisher = {IEEE},
  doi       = {10.23919/IFIPNetworking.2018.8696555}
}

@inproceedings{Demers1987Epidemic,
  author       = {Alan Demers and Dan Greene and Carl Hauser and Wes Irish and John Larson and Scott Shenker and Howard Sturgis and Dan Swinehart and Doug Terry},
  title        = {Epidemic Algorithms for Replicated Database Maintenance},
  booktitle    = {Proceedings of the Sixth Annual ACM Symposium on Principles of Distributed Computing},
  pages        = {1--12},
  year         = {1987},
  publisher    = {ACM},
  doi          = {10.1145/41840.41841},
  url          = {https://doi.org/10.1145/41840.41841}
}

@inproceedings{Douceur2002Sybil,
  author       = {John R. Douceur},
  title        = {The Sybil Attack},
  booktitle    = {Peer-to-Peer Systems: First International Workshop, IPTPS 2002},
  year         = {2002},
  pages        = {251--260},
  publisher    = {Springer},
  doi          = {10.1007/3-540-45748-8_24}
}

@misc{Ethereum2026Pos,
  author       = {{Ethereum Foundation}},
  title        = {Proof-of-Stake Rewards and Penalties},
  year         = {2026},
  url          = {https://ethereum.org/developers/docs/consensus-mechanisms/pos/rewards-and-penalties/},
  note         = {Accessed: 2026-05-06}
}

@misc{Europeancommission2026Eudi,
  author       = {{European Commission}},
  title        = {EU Digital Identity Wallet},
  year         = {2026},
  url          = {https://ec.europa.eu/digital-building-blocks/sites/spaces/EUDIGITALIDENTITYWALLET/pages/694487738/EU+Digital+Identity+Wallet+Home},
  note         = {Accessed: 2026-05-06}
}

@article{FilippiLoveluck2016Bitcoinpolitics,
  author       = {Primavera De Filippi and Benjamin Loveluck},
  title        = {The Invisible Politics of Bitcoin: Governance Crisis of a Decentralised Infrastructure},
  journal      = {Internet Policy Review},
  volume       = {5},
  number       = {3},
  year         = {2016},
  doi          = {10.14763/2016.3.427}
}

@inproceedings{FokkerPouwelseBuntine2006P2PWikipedia,
  author    = {Fokker, Jenneke E. and Pouwelse, Johan A. and Buntine, Wray},
  title     = {{Tag-Based Navigation for Peer-to-Peer Wikipedia}},
  booktitle = {{Collaborative Web Tagging Workshop at WWW2006}},
  address   = {Edinburgh, United Kingdom},
  year      = {2006}
}

@article{FritschMullerWattenhofer2022DAOControl,
  author        = {Robin Fritsch and Marino M{\"u}ller and Roger Wattenhofer},
  title         = {{Analyzing Voting Power in Decentralized Governance: Who Controls DAOs?}},
  journal       = {arXiv preprint arXiv:2204.01176},
  year          = {2022},
  eprint        = {2204.01176},
  archivePrefix = {arXiv},
  primaryClass  = {cs.CY},
  url           = {https://arxiv.org/abs/2204.01176}
}

@misc{GalaxyBitcoinFees2025,
  author       = {{Galaxy Research}},
  title        = {{Bitcoin Fees Collapse: What Onchain Data Tells Us}},
  howpublished = {\url{https://www.galaxy.com/insights/research/bitcoin-onchain-fees-utxo}},
  year         = {2025},
  note         = {Accessed 2026-06-20}
}

@misc{Gitcoin2026QuadraticFunding,
  author       = {{Gitcoin}},
  title        = {{Quadratic Funding}},
  year         = {2026},
  url          = {https://gitcoin.co/mechanisms/quadratic-funding},
  note         = {Accessed: 2026-05-31}
}

@misc{GitHub2026AvailabilityUpdate,
  author       = {Fedorov, Vlad},
  title        = {{An update on GitHub availability}},
  year         = {2026},
  month        = apr,
  day          = {28},
  url          = {https://github.blog/news-insights/company-news/an-update-on-github-availability/},
  note         = {Accessed: 2026-06-09}
}

@misc{GitHub2026CopilotUsageBilling,
  author       = {{GitHub}},
  title        = {{GitHub Copilot is moving to usage-based billing}},
  year         = {2026},
  month        = apr,
  url          = {https://github.blog/news-insights/company-news/github-copilot-is-moving-to-usage-based-billing/},
  note         = {Accessed: 2026-06-08}
}

@misc{GitHub2026CopilotLimits,
  author       = {{GitHub}},
  title        = {{Changes to GitHub Copilot Individual plans}},
  year         = {2026},
  month        = apr,
  url          = {https://github.blog/news-insights/company-news/changes-to-github-copilot-individual-plans/},
  note         = {Accessed: 2026-06-08}
}

@inproceedings{HalesRahmanZhangMeulpolderPouwelse2009BitCrunch,
  author       = {Hales, David and Rahman, Rameez and Zhang, Boxun and Meulpolder, Michel and Pouwelse, Johan},
  title        = {{BitTorrent or BitCrunch: Evidence of a Credit Squeeze in BitTorrent?}},
  booktitle    = {{2009 18th IEEE International Workshops on Enabling Technologies: Infrastructures for Collaborative Enterprises}},
  pages        = {99--104},
  year         = {2009},
  publisher    = {IEEE},
  doi          = {10.1109/WETICE.2009.22}
}

@misc{Humanpassport2026Docs,
  author       = {{Human Passport}},
  title        = {Human Passport Developer Platform},
  year         = {2026},
  url          = {https://docs.passport.xyz/},
  note         = {Accessed: 2026-05-06}
}

@misc{InvestingBitcoinHistoricalData,
  author       = {{Investing.com}},
  title        = {{BTC/USD Bitfinex Historical Data}},
  howpublished = {\url{https://www.investing.com/crypto/bitcoin/btc-usd-historical-data}},
  year         = {2026},
  note         = {Accessed 2026-06-20}
}

@misc{JozwikPouwelse2025SmartphoneDemocracy,
  title        = {SmartphoneDemocracy: Privacy‐Preserving E‐Voting on Decentralized Infrastructure using Novel European Identity},
  author       = {Micha{\l} Jo{\'z}wik and Johan Pouwelse},
  year         = {2025},
  url          = {https://arxiv.org/abs/2507.09453},
  eprint       = {2507.09453},
  archivePrefix= {arXiv},
  primaryClass = {cs.CR}
}

@inproceedings{KarpSchindelhauerShenkerVocking2000RumorSpreading,
  author       = {Richard Karp and Christian Schindelhauer and Scott Shenker and Berthold V{\"o}cking},
  title        = {Randomized Rumor Spreading},
  booktitle    = {Proceedings of the 41st Annual Symposium on Foundations of Computer Science},
  pages        = {565--574},
  year         = {2000},
  publisher    = {IEEE},
  doi          = {10.1109/SFCS.2000.892324},
  url          = {https://doi.org/10.1109/SFCS.2000.892324}
}

@misc{Libp2pGossipsubV11,
  author       = {{libp2p}},
  title        = {{GossipSub v1.1: Security Extensions to Improve on Attack Resilience and Bootstrapping}},
  url          = {https://github.com/libp2p/specs/blob/master/pubsub/gossipsub/gossipsub-v1.1.md},
  note         = {Accessed: 2026-05-28}
}

@article{LiXuDuan2023LiquidDPoS,
  author        = {Chao Li and Runhua Xu and Li Duan},
  title         = {{Liquid Democracy in DPoS Blockchains}},
  journal       = {arXiv preprint arXiv:2309.01090},
  year          = {2023},
  eprint        = {2309.01090},
  archivePrefix = {arXiv},
  primaryClass  = {cs.CY},
  url           = {https://arxiv.org/abs/2309.01090}
}

@mastersthesis{DogariuPouwelse2026SelfReplicatingSeedboxes,
  author  = {Matei Dogariu and Johan Pouwelse},
  title   = {{Self-replicating seedbox servers using programmable money}},
  school  = {Delft University of Technology},
  type    = {Master's thesis},
  year    = {2026},
  address = {Delft, The Netherlands},
  note    = {Forthcoming in the TU Delft Repository},
  url     = {https://repository.tudelft.nl/}
}

@article{MockusFieldingHerbsleb2002Oss,
  author       = {Audris Mockus and Roy T. Fielding and James D. Herbsleb},
  title        = {Two Case Studies of Open Source Software Development: Apache and Mozilla},
  journal      = {ACM Transactions on Software Engineering and Methodology},
  publisher    = {Association for Computing Machinery},
  address      = {New York, NY, USA},
  volume       = {11},
  number       = {3},
  pages        = {309--346},
  year         = {2002},
  issue_date   = {July 2002},
  issn         = {1049-331X},
  doi          = {10.1145/567793.567795}
}

@misc{Nakamoto2008Bitcoin,
  title        = {Bitcoin: A Peer-to-Peer Electronic Cash System},
  author       = {Satoshi Nakamoto},
  year         = {2008},
  url          = {https://bitcoin.org/bitcoin.pdf},
  note         = {Accessed: 2026-05-06}
}

@techreport{NIST2023DSS,
  title        = {{Digital Signature Standard (DSS)}},
  institution  = {National Institute of Standards and Technology},
  type         = {Federal Information Processing Standards Publication},
  number       = {FIPS 186-5},
  year         = {2023},
  month        = feb,
  doi          = {10.6028/NIST.FIPS.186-5},
  url          = {https://doi.org/10.6028/NIST.FIPS.186-5}
}

@techreport{NIST2025DigitalIdentity,
  title        = {{Digital Identity Guidelines: Authentication and Authenticator Management}},
  institution  = {National Institute of Standards and Technology},
  type         = {Special Publication},
  number       = {NIST SP 800-63B-4},
  year         = {2025},
  url          = {https://pages.nist.gov/800-63-4/sp800-63b.html}
}

@article{PeltJansenBaarsOverbeek2020Governance,
  author       = {Rowan van Pelt and Slinger Jansen and Djuri Baars and Sietse Overbeek},
  title        = {Defining Blockchain Governance: A Framework for Analysis and Comparison},
  journal      = {Information Systems Management},
  volume       = {38},
  number       = {1},
  pages        = {21--41},
  year         = {2021},
  publisher    = {Taylor \& Francis},
  doi          = {10.1080/10580530.2020.1720046}
}

@misc{RFC5694PeerToPeer,
  author       = {Gonzalo Camarillo},
  title        = {{Peer-to-Peer (P2P) Architecture: Definition, Taxonomies, Examples, and Applicability}},
  howpublished = {RFC 5694},
  year         = {2009},
  month        = nov,
  publisher    = {RFC Editor},
  doi          = {10.17487/RFC5694},
  url          = {https://www.rfc-editor.org/rfc/rfc5694}
}

@misc{RFC9380HashToCurve,
  author       = {Armando Faz-Hernandez and Sam Scott and Nick Sullivan and Riad S. Wahby and Christopher A. Wood},
  title        = {{Hashing to Elliptic Curves}},
  howpublished = {RFC 9380},
  year         = {2023},
  month        = aug,
  doi          = {10.17487/RFC9380},
  url          = {https://www.rfc-editor.org/rfc/rfc9380}
}

@inproceedings{Schollmeier2001PeerToPeer,
  author       = {R{\"u}diger Schollmeier},
  title        = {{A Definition of Peer-to-Peer Networking for the Classification of Peer-to-Peer Architectures and Applications}},
  booktitle    = {Proceedings of the First International Conference on Peer-to-Peer Computing},
  pages        = {101--102},
  year         = {2001},
  publisher    = {IEEE},
  doi          = {10.1109/P2P.2001.990434},
  url          = {https://doi.org/10.1109/P2P.2001.990434}
}

@article{SiddarthIvlievSiriBerman2020Watchmen,
  title        = {Who Watches the Watchmen? A Review of Subjective Approaches for Sybil-Resistance in Proof of Personhood Protocols},
  author       = {Divya Siddarth and Sergey Ivliev and Santiago Siri and Paula Berman},
  journal      = {Frontiers in Blockchain},
  volume       = {Volume 3 - 2020},
  year         = {2020},
  url          = {https://www.frontiersin.org/journals/blockchain/articles/10.3389/fbloc.2020.590171},
  doi          = {10.3389/fbloc.2020.590171},
  issn         = {2624-7852},
}

@misc{TezosGovernance,
  author       = {{Tezos}},
  title        = {{Governance and Self-Amendment}},
  year         = {2025},
  url          = {https://docs.tezos.com/architecture/governance}
}

@misc{VyzovitisNaporaMcCormickDiasPsaras2020GossipSub,
  author       = {Dimitris Vyzovitis and Yusef Napora and Dirk McCormick and David Dias and Yiannis Psaras},
  title        = {{GossipSub: Attack-Resilient Message Propagation in the Filecoin and ETH2.0 Networks}},
  year         = {2020},
  eprint       = {2007.02754},
  primaryClass = {cs.NI},
  url          = {https://arxiv.org/abs/2007.02754}
}

@article{Zhou2021BountiesGitHub,
  author  = {Zhou, Jiayuan and Wang, Shaowei and Bezemer, Cor-Paul and Zou, Ying and Hassan, Ahmed E.},
  title   = {{Studying the Association Between Bountysource Bounties and the Issue-Addressing Likelihood of GitHub Issue Reports}},
  journal = {IEEE Transactions on Software Engineering},
  volume  = {47},
  number  = {12},
  pages   = {2919--2933},
  year    = {2021},
  doi     = {10.1109/TSE.2020.2974469}
}

@misc{ZichichiContuFerrettiDAngelo2019LikeStarter,
  author       = {Zichichi, Mirko and Contu, Michele and Ferretti, Stefano and D'Angelo, Gabriele},
  title        = {{LikeStarter: A Smart-contract based Social DAO for Crowdfunding}},
  year         = {2019},
  eprint       = {1905.05560},
  archivePrefix = {arXiv},
  primaryClass = {cs.CY},
  url          = {https://arxiv.org/abs/1905.05560}
}

\newpage
\appendix

\section{Acknowledgement on the use of AI}

Throughout the research, we used AI tools to challenge, improve, and simplify parts of our work. In particular, we used ChatGPT 5\footnote{\href{https://chatgpt.com}{chatgpt.com}} as the central model for most of our questions and revisions. The following subsections describe how ChatGPT and other special-purpose AI tools were used during the different phases of the research.

\subsection{Literature research and design}
At the start of the research, we intentionally limited the use of AI tools. This allowed us to explore the topic without letting AI-generated suggestions determine the initial direction of the work. Instead, we first focused on developing our own understanding of the problem space and identifying the main research direction independently.

Once the research direction became clearer and concrete design ideas began to form, we incorporated AI tools into the design process. During this phase, we used ChatGPT as a sparring partner to explain, test, and challenge our ideas. We also used it to suggest alternative design options, which helped us compare possible approaches before making design decisions.

\subsection{Coding}
We used generative AI tools extensively during the implementation phase. ChatGPT was often used to discuss architecture and draft initial code versions. After reviewing these drafts, we frequently used Codex\footnote{\href{https://openai.com/codex}{openai.com/codex}} to improve and simplify the code. All generated code was reviewed before being committed.

The implemented code was tested where possible. For testing, we used AI tools to draft initial test cases. However, to avoid relying on tests that merely confirmed the generated implementation, we carefully checked the inputs and expected outputs ourselves. When required, these tests were modified, and missing test cases were added manually.

The front-end design was important for demonstrating the system, but it was not the main focus of the research. For this reason, we used Stitch\footnote{\href{https://stitch.withgoogle.com}{stitch.withgoogle.com}} to quickly create a draft user interface. The generated design, both as an image and as HTML/CSS, was then used as input for ChatGPT, which helped convert it into Qt code. This code was subsequently refined using Codex.

\subsection{Writing}
When it comes to writing, we utilised AI tools solely to ensure consistency in style, grammar, and spelling. Specifically, we employed ChatGPT to rephrase sections of the document to maintain a unified tone. The introduction of the report served as our style reference. As new sections were drafted, we used ChatGPT to align these with the preferred style. However, rather than accepting the AI-generated output immediately, we carefully selected and refined the parts that aligned with our vision, ensuring the final product remained clear, avoiding any overly abstract phrasing that can sometimes arise from generative AI tools.

In addition to ChatGPT, we used the pro version of Grammarly\footnote{\href{https://grammarly.com}{grammarly.com}} to catch grammar and spelling errors. While Grammarly also suggested improvements for clarity, engagement, and delivery, we applied those recommendations only when they aligned with our intended message and tone.

\subsection{Responsibility, verification, and data handling}
All AI-generated suggestions, text, code, and design alternatives were critically reviewed before being included in the paper or implementation. We remain fully responsible for the final content, including the accuracy, originality, and integrity of the research, arguments, code, and written text. AI tools were not treated as authoritative sources. Any factual claims, background information, or ideas derived from external literature were verified against the original sources and cited accordingly.

\end{document}